\documentclass[11pt]{article}
\usepackage{srcltx,amsmath,epsfig,sint}
\usepackage{pst-all}
\usepackage{slashed}
\usepackage{latexsym}
\usepackage{multirow}
\usepackage[american]{babel}

\newcommand{\unit}{1\kern-.25em {\rm l}}

\newcommand{\be}{\begin{equation}}
\newcommand{\ee}{\end{equation}}
\newcommand{\bd}{\begin{displaymath}}
\newcommand{\ed}{\end{displaymath}}
\newcommand{\bea}{\begin{eqnarray}}
\newcommand{\eea}{\end{eqnarray}}
\newcommand{\ba}{\begin{array}}
\newcommand{\ea}{\end{array}}

\newcommand{\tr}{{\rm tr}}

\newcommand{\psibar}{\bar{\psi}}

\newcommand{\Real}{\relax{\mathsf{\Gamma\kern-.35em R}}}
\newcommand{\Int}{\relax{\mathsf{Z\kern-.40em Z}}}

\newcommand{\diag}{{\rm diag}}

\newcommand{\Nf}{N_{\rm f}}

\newcommand{\bfx}{{\bf x}}

\newcommand{\rme}{{\rm e}}
\newcommand{\rmO}{{\rm O}}

\newcommand{\zetabar}{\bar{\zeta}}
\newcommand{\zetaprime}{\zeta\kern1pt'}
\newcommand{\zetabarprime}{\zetabar\kern1pt'}

\newcommand{\msbar}{{\rm \overline{MS\kern-0.05em}\kern0.05em}}

\def\cs{c_{\rm s}}
\def\ct{c_{\rm t}}

\def\rmd{{\rm d}}

\begin{document}
  \begin{titlepage}
\begin{flushright}
KANAZAWA-11-09\\
April 2011\\
\end{flushright} 

\vskip 1 cm
\begin{center}
  {\Large\bf 
An O($a$) modified lattice set-up of the Schr\"odinger functional \\[1ex]
in SU(3) gauge theory}
\end{center}
\vskip 3ex

\begin{figure}[h]
\begin{center}
\epsfig{figure=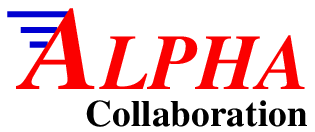}
\end{center}
\end{figure}

\begin{center}
{\large Paula P\'erez Rubio$^a$, Stefan Sint$^b$ and Shinji Takeda$^c$}
\vskip 2.3ex
\begin{flushleft}
$^a$ Institut f\"ur Theoretische Physik, Universit\"at Regensburg, 93040 Regensburg, Germany\\[1ex]
$^b$ School of Mathematics, Trinity College Dublin, Dublin 2, Ireland\\[1ex]
$^c$ School of Mathematics and Physics, College of Science and Engineering, \\
\hspace{2ex} Kanazawa University, Kakuma-machi, Kanazawa, Ishikawa 920-1192, Japan\\[1ex]
\end{flushleft}
\vskip 1ex
{\bf Abstract}
\vskip 0.7ex
\end{center}
The set-up of the QCD Schr\"odinger functional (SF) on the lattice with staggered quarks 
requires an even  number of points $L/a$ in  the spatial directions, 
while the Euclidean time extent of the lattice, $T/a$, must be odd. 
Identifying a unique renormalisation scale, $L=T$, is then 
only possible up to O($a$) lattice artefacts. 
In this article we study such lattices in the pure SU(3) gauge theory, 
where we can also compare to the standard set-up. We consider the 
SF coupling as obtained from the variation of an 
SU(3) Abelian and spatially constant background field. The O($a$) lattice 
artefacts can be cancelled by the existing O($a$) boundary counterterm. 
However, its coefficient, $\ct$, differs at the tree-level from its standard value, 
so that one first needs to re-determine the induced background gauge field.
The perturbative one-loop correction to the coupling allows to determine $\ct$ 
to one-loop order. A few numerical simulations serve to demonstrate 
that residual cutoff effects in the step scaling function are small in both cases, 
$T=L\pm a$ and comparable to the standard case with $T=L$.
\vfill
\eject
\end{titlepage}

  \section{Introduction}

Renormalisation schemes based on the Schr\"odinger functional \cite{Luscher:1992an-Sint:1995rb} 
have come to play an important r\^ole in lattice QCD and in Technicolor
inspired models of electroweak symmetry breaking. The Schr\"odinger functional (SF) is used 
to define a technically convenient, intermediate renormalisation  scheme, where the scale 
is set by the finite space-time volume, and all fermion masses are taken to vanish. In SF schemes, 
the non-perturbative scale evolution of the running coupling or multiplicative renormalisation 
constants of composite operators can then be constructed recursively 
(see \cite{Jansen:1995ck,Sommer:2006sj} for an introduction). 
Note that the scale evolution is eventually obtained in the continuum limit and thus universal.
Hence, the choice of the lattice regularisation for this part of the calculation 
is merely a matter of practical considerations. In particular, in order to minimise computational
costs it is advisable to use some variant of either Wilson or staggered quarks.

The Schr\"odinger functional in lattice QCD was originally obtained for 
Wilson type quarks~\cite{Sint:1993un}. Its formulation for staggered/Kogut-Susskind quarks has been initiated
in \cite{Miyazaki:1994nu} and further studied in~\cite{Heller:1997pn}. 
For theories with a multiple of four massless fermion 
flavours\footnote{For staggered quarks we here use the traditional term 
``flavour" rather than ``taste''.},  staggered quarks constitute  an attractive alternative 
to Wilson quarks: no tuning is required to recover the chiral limit, and numerical simulations for a given
lattice size are computationally less demanding. In fact, the SF coupling has already been studied 
with $\Nf=16$~\cite{Heller:1997vh} and $\Nf=8,12$ \cite{Appelquist:2007hu} fermion flavours,
and a four-flavour QCD study is in preparation~\cite{PerezRubio:PhD-PerezRubio:prep1}.
A drawback, however, consists in the observation that staggered fermions require lattices
with an even extent $L/a$ in the spatial directions, 
whereas the time extent $T/a$ must be odd~\cite{Miyazaki:1994nu,Heller:1997pn}. 
As the renormalisation scale must be given by a unique scale,  
$\mu=L^{-1}$, the ratio $T/L$ must be fixed and is usually set to $T/L=1$. 
With staggered quarks,  this can only be achieved up to corrections of O($a$)
and the question arises how the ensuing O($a$) effects can be eliminated. For the running coupling, 
Heller proposed to take an average of observables on lattices with $T=L\pm a$, 
and he showed that this  procedure is consistent to one-loop order of perturbation theory. 
The same recipe was then also applied in numerical simulations~\cite{Heller:1997vh,Appelquist:2007hu}. 
However, it is not obvious how this procedure can be generalised to fermionic 
correlation functions~\cite{PerezRubio:2008yd}. 

It is thus desirable to find an alternative to the averaging procedure. It is the purpose
of the present paper to show how this is possible by slightly modifying the approach to the continuum limit.
More precisely, the limit should be taken at fixed $T'/L$ where $T'$ 
is either set to $T+a$ or to $T-a$. Requiring O($a$) improvement of the pure gauge theory leads to 
modifications of the boundary O($a$) counterterm proportional to $\ct$ already at the tree-level. 
This in turn affects the equations of motion for the gauge field and thus changes 
the minimum action configuration used to define the coupling in the SF scheme.
We  determine the new background gauge field and then perform a one-loop computation 
to calculate the boundary O($a$) improvement of the SF to this order for the pure gauge theory. 
A numerical simulation has been carried out to check for the size of residual cutoff effects in the 
step-scaling function for the SF coupling.
Although our motivation for this work originates in the fermionic sector of the SF,
we here only discuss the necessary modifications to the SF in the pure gauge theory~\cite{Luscher:1992an}.
The details and calculations pertaining to the fermionic sector with staggered quarks will be presented
elsewhere~\cite{PerezRubio:PhD,PerezRubio:prep2}. 

This paper is organised as follows: after a brief review of the continuum Schr\"odinger functional and the 
definition of the renormalised coupling, we discuss its lattice regularization and the
origin of the constraint on lattice sizes with staggered fermions (Sect.~2).
In Sect.~3, the background gauge field is found with $\ct$ left as parameter, which is then determined
such as to cancel O($a$) effects in the action.  We then proceed with the one-loop calculation of the running coupling (Sect.~4), 
and determine $\ct$ to this order, as well as the size of remaining lattice artefacts at one-loop order.
In Sect.~5 we discuss the results of our numerical simulation at an intermediate value of the running coupling. 
Conclusions are presented in Sect.~6. For future reference, 
an appendix contains two tables with the raw data
of our perturbative one-loop calculation.

  \section{A short reminder of the Schr\"odinger functional coupling}

In this section we review some known facts and definitions which will be relevant 
for this paper. In the continuum the Schr\"odinger functional is formally defined as the
Euclidean path integral,
\begin{equation}
   {\cal Z}[C,C']= \int D[A,\psi,\psibar]{\rm e}^{-S_{\rm cont}[A,\psi,\psibar]},
\end{equation}
with the Euclidean action 
\begin{eqnarray}
   S_{\rm cont} &=& \int_0^T\rmd x_0 \int_0^L\rmd^3{\bfx}\, {\cal L}(x),\\
   {\cal L}(x) &=& -\frac{1}{2g_0^2}\tr\{F_{\mu\nu}(x) F_{\mu\nu}(x)\} 
   + \psibar(x)(\gamma_\mu D_\mu+m)\psi(x).
\end{eqnarray}
Here, $g_0$ denotes the bare coupling constant, $F_{\mu\nu}$ is the field tensor 
associated with the gauge field $A_\mu$,
\begin{equation}
  F_{\mu\nu}=\partial_\mu A_\nu-\partial_\nu A_\mu +[A_\mu,A_\nu],
\end{equation}
and $D_\mu=\partial_\mu+A_\mu+i\theta_\mu/L$ denotes the covariant derivative acting on
the quark fields (including a constant U(1) background field with $\theta_\mu= (1-\delta_{\mu 0})\theta$).
The space-time manifold is taken to be a hyper-cylinder. In the spatial directions 
periodic boundary conditions are imposed on all fields. At the time boundaries 
the conditions for the fermionic fields read
\begin{equation}
  P_+\psi\vert_{x_0=0}=0=P_-\psi\vert_{x_0=T},\qquad
  \psibar P_-\vert_{x_0=0} =0= \psibar P_+\vert_{x_0=T},
\end{equation}
with the projectors $P_\pm =\frac12(1\pm\gamma_0)$.
For the gauge field one has
\begin{equation}
   A_k|_{x_0=0}  =  C^{}_k,\qquad
   A_k|_{x_0=T}  =   C'_k,\qquad k=1,2,3,
   \label{eq:bcs_gaugefields}
\end{equation}
with the boundary gauge field configurations $C_k$ and $C_k'$.
Of particular interest are spatially constant Abelian fields, which take the form,
\begin{equation}
  C^{}_k=\frac{i}{L}\phi, \qquad
  C'_k  =\frac{i}{L}\phi',  \qquad k=1,2,3,
 \label{eq:bc_phases}
\end{equation}
with traceless and diagonal $N\times N$-matrices $\phi$ and $\phi'$.
For $N=3$ colours we follow~\cite{Luscher:1993gh} and parameterise the diagonal elements by,
\begin{xalignat}{2}
  \phi^{}_1 & = \eta-\frac{\pi}{3}, & 
    & \phi'_1 = -\eta-\pi,  \nonumber\\
  \phi^{}_2 & = \eta\left(\nu-\frac12\right),  & 
    & \phi'_2 =  \eta\left(\nu+\frac12\right)+\frac{\pi}{3},    \label{eq:phi_phiprime}\\
  \phi^{}_3 & = -\eta\left(\nu+\frac12\right)+\frac{\pi}{3}, & 
    &\phi'_3  = -\eta\left(\nu-\frac12\right)+\frac{2\pi}{3},   \nonumber   
\end{xalignat}
where $\eta$ and $\nu$ are 2 real parameters.
In the temporal gauge the field equations with these boundary conditions are solved by,
\begin{equation}
  B_0=0,\qquad B_k= C_k + \frac{x_0}{T}\left(C_k'-C_k\right),  \qquad k=1,2,3. 
  \label{eq:contBF}
\end{equation}
which corresponds to a constant chromo-electric field,
\begin{equation}
  G_{0k}=\partial_0 B_k = \frac{C_k'-C_k}{T}= \frac{i(\phi'-\phi)}{LT}, \qquad k=1,2,3,
\end{equation}
all chromo-magnetic components of the field tensor being zero.
The background field $B$ is, up to gauge transformations, uniquely determined by the
gauge boundary fields since it corresponds to the absolute minimum of the gauge action,
\begin{equation}
  S_{\rm cont}[B]= \frac{3}{\rho g_0^2}\sum_{\alpha=1}^3(\phi'_\alpha-\phi^{}_\alpha)^2 = 
  \frac{18}{\rho g_0^2}\left(\eta+\frac{\pi}{3}\right)^2,
\label{eq:Scont}
\end{equation}
where $\rho=T/L$ is the aspect ratio.
The effective action for the SF can thus be taken to be a function of
the background field,
\begin{equation}
  \Gamma[B]=-\ln {\cal Z}[C',C],
\end{equation}
which admits a weak coupling expansion,
\begin{equation}
  \Gamma[B] \, \buildrel g_0\rightarrow0\over\sim\,\, \frac{1}{g_0^2}\Gamma_0[B]+ \Gamma_1[B]+\rmO(g_0^2),
\label{eq:coup_exp}
\end{equation}
with the lowest order term given by the classical action, $\Gamma_0[B]=g_0^2 S_{\rm cont}[B]$.
Setting  $T=L$, one remains with a single external scale, $L$,
and a renormalised coupling $\bar{g}(L)$ can now be defined through
\begin{equation}
  {\frac{\partial\Gamma}{\partial\eta}}
  \biggl\vert_{\eta=0}= k\left\{\frac{1}{{\bar g}^2(L)}-\nu \bar v(L)\right\},   
  \qquad k={\frac{\partial\Gamma_0}{\partial\eta}}
  \biggl\vert_{\eta=0}=12\pi.
  \label{eq:coupling}
\end{equation}
Here the $\eta$-derivative serves to eliminate any background field independent terms 
in the effective action. Moreover, it implies that the coupling is defined by an 
SF correlation function rather than by the SF itself, which renders 
it measurable by numerical simulations. In practice the parameter $\nu$ will be set
to zero. However, its coefficient in Eq.~(\ref{eq:coupling}), $\bar{v}$, defines a further
observable, which can also be measured at $\nu=0$, based on the relation,
\begin{equation}
 \bar{v}(L) = -\frac{1}{k} \frac{\partial}{\partial\nu}\left\{\left.
   \frac{\partial\Gamma}{\partial\eta}\right\vert_{\eta=0}\right\}_{\nu=0}.
\label{eq:v_def}
\end{equation}
In numerical simulations, the scale evolution for the coupling is constructed non-perturbatively
by computing the continuum step-scaling function
\begin{equation}
   \sigma(u)={\bar g}^2(2L)\vert_{u={\bar g}^2(L)}.
\end{equation}
Prescribing a value for  $u={\bar g}^2(L)$ implicitly fixes the scale $L$, and thus the
only free parameter in pure Yang-Mills theory. The coupling at the scale $2L$ or $\bar{v}(L)$ are
then fixed and can be obtained by taking the continuum limit of the corresponding observables on the lattice.
We will later use the deviation from their respective continuum limits $\sigma(u)$ and 
\begin{equation}
   \omega(u)={\bar v}(L)\vert_{u={\bar g}^2(L)},
  \label{eq:omega}
\end{equation}
to obtain an impression of the cutoff effects for the different lattice regularisations considered here.

The SF coupling is gauge invariant and can be non-perturbatively defined on the lattice.
For small volumes it can be related perturbatively to the $\overline{\rm MS}$
scheme. This relation is known to two-loop order both for the pure SU(3) gauge theory~\cite{Bode:1998hd}
and QCD~\cite{Bode:1999sm}. Setting $q=1/L$ and $\alpha=\bar{g}^2/4\pi$, the 
result for pure SU(3) gauge theory is
\begin{equation}
  \alpha_{\overline{\rm MS}}(q)=\alpha(q)+c_1\alpha^2(q)+c_2\alpha^3(q)+\rmO(\alpha^4),
\label{eq:c1}
\end{equation}
with~\cite{Bode:1998hd},
\begin{equation}
   c_1= 1.255621(2),\qquad c_2=c_1^2 + 1.197(10).
\end{equation}
For the perturbative checks in this paper we will also need the 
one-loop result for the observable $\bar{v}(L)$~\cite{Luscher:1993gh},
\begin{equation}
  \omega(u)= \omega_1 + \omega_2 u + \rmO(u^2),
  \qquad \omega_1= 0.0694603(1).
  \label{eq:vbar1loop}
\end{equation}
Finally, we recall that the step-scaling function to first non-trivial order is given by
\begin{equation}
   \sigma(u)= u+ 2b_0\ln(2)\times u^2 +\rmO(u^3), \qquad 
    b_0= 11/(4\pi)^{2}.  
    \label{eq:SU3beta}
\end{equation}
Here, $b_0$ is the universal one-loop coefficient of the SU(3) $\beta$-function,
which, for the SF coupling is known to 3-loop order~\cite{Bode:1998hd}.

  \section{Variants of the pure gauge Schr\"odinger functional on the lattice}

We here summarise the basic definitions and properties of the SF in the SU($N$) lattice gauge theory. 
The change at O($a$) motivated by staggered fermions has tree-level consequences
which will be worked out in the remainder of this section.

\subsection{The standard lattice framework}

The lattice formulation of the SF for pure SU($N$) gauge theories can be obtained from the transfer 
matrix formalism~\cite{Luscher:1992an}. When written as a path integral it takes the form
\begin{equation}
 {\cal Z}[C,C'] = \int D[U] \rme^{-S[U]}.
\end{equation}
The pure gauge action is given by
\begin{equation}
  S[U] = \frac{1}{g_0^2}\sum_p w(p) {\rm tr}\{1 - U(p)\},
  \label{eq:Slatt} 
\end{equation}
where the sum runs over all oriented plaquettes $p$ of the lattice and $U(p)$ 
denotes the parallel transporter around $p$. Assuming Abelian boundary gauge fields $C_k$ and $C'_k$,
the boundary conditions, Eq.~(\ref{eq:bcs_gaugefields}) translate to
\begin{equation}
    U(x,k)\vert_{x_0=0}= W(\bfx,k),\qquad U(x,k)\vert_{x_0=T}= W'(\bfx,k),
\end{equation}
with the boundary link variables,
\begin{equation}
   W(\bfx,k)=\exp{\left(aC_k^{}\right)},\qquad  
   W'(\bfx,k)=\exp{\left(aC'_k\right)}.
\end{equation}
Finally, the weight factor, $w(p)$, is set to
\begin{equation}
w(p) = \left\{\begin{array}{ll} 
   \frac 12 \cs(g_0) & \text{for spatial plaquettes $p$ at } x_0= 0, T,\\
            \ct(g_0) & \text{for time-like plaquettes $p$ attached to the boundaries,}\\ 
                   1 & \text{otherwise.}
             \end{array}
        \right.
\end{equation}
In the continuum limit the plaquette terms multiplied by $\ct$ and $\cs$ respectively reduce to the 
only gauge invariant local boundary operators of dimension 4, which are allowed by the symmetries of the SF
(repeated spatial Lorentz indices are summed over),
\begin{equation}
   \tr\{F_{0k}F_{0k}\},\qquad \tr\{F_{kl}F_{kl}\}.
\end{equation}
There are no other sources for lattice artefacts linear in $a$, and 
one may thus cancel O($a$) effects in any on-shell quantity by properly adjusting these coefficients.
Moreover, while $\ct$ contributes to all observables, the counterterm proportional to $\cs$ vanishes
for spatially constant Abelian boundary gauge fields. In this case,
O($a$) improvement can be achieved with the appropriate choice of $\ct$ alone.
In perturbation theory, one has
\begin{equation}
   \ct(g_0) = \ct^{(0)} + \ct^{(1)} g_0^2 + \dots 
\end{equation}
In the standard set-up of the SF one sets $L=T$ and tree-level O($a$) improvement
is achieved by setting $\ct=1$. With this choice the classical equations of motion
which follow from the lattice action are equivalent to 
\begin{equation}
   d^\ast P(x,\mu) - d^\ast P(x,\mu)^\dagger 
   -\frac{1}{N}\tr\left\{ d^\ast P(x,\mu) - d^\ast P(x,\mu)^\dagger \right\} =0,
\label{eq:eom}
\end{equation}
where
\begin{equation}
  d^\ast P(x,\mu) = \sum_{\nu=0}^3\left\{P_{\mu\nu}(x) 
    - U_\nu^\dagger(x-a \hat\nu) P_{\mu\nu}(x-a\hat\nu)U_\nu^{}(x-a\hat\nu)\right\},
 \label{eq:covdiv}
\end{equation}
and the plaquette field is defined by
\begin{equation}
    P_{\mu\nu} (x) = U_\mu(x) U_\nu(x + a\hat \mu) U_\mu^\dagger(x + a\hat\nu)U_\nu^\dagger(x).
\end{equation}
A solution $V_\mu(x)$ to Eq.~(\ref{eq:eom}) which is unique up to gauge equivalence, 
is referred to as the lattice background gauge field. 
For spatially constant Abelian boundary fields, the lattice background field is obtained
by simply exponentiating the continuum solution, Eq.~(\ref{eq:contBF}),
\begin{equation}
   V_\mu(x)=\exp(aB_\mu(x)), \qquad \mu=0,\ldots,3.
\end{equation}
Moreover, a mathematical proof establishes that this solution corresponds
to an absolute minimum of the action, provided the condition 
\begin{equation}
   TL/a^2>(N-1)\pi^2{\rm max}\{1,N/16\},
\end{equation}
is met~\cite{Luscher:1992an}. For $N=3$ colours, this means that $S[V]$ represents an absolute minimum for 
$TL/a^2\geq 2\pi^2$, implying $L/a>4$ on lattices with $L=T$.
The action at the minimum,
\begin{equation}
  S[V]=\frac{3TL^3}{g_0^2}\sum_{\alpha=1}^3
  \left\{\frac{2}{a^2}\sin\left[\frac{a^2}{2TL}(\phi_\alpha'-\phi_\alpha)\right] \right\}^2,
\end{equation}
only differs at O($a^4$) from the continuum action, Eq.~(\ref{eq:Scont}).

One should note, however, that these considerations depend on having set $\ct=1$. 
As we will now explain, staggered fermions require a modified approach to the continuum limit,
which will lead to a different choice of $\ct$. Consequently, the determination of the
classical background field needs to be revisited.

\subsection{A constraint from staggered quarks}

\begin{figure}[ht!]
\begin{center}
\begin{center}
\scalebox{1.0} 
{

\begin{pspicture}(0,-2.485)(11.34,2.485)
\definecolor{color38}{rgb}{0.17,0.56,0.027}
\definecolor{color41}{rgb}{0.5,0.5,0.5}
\definecolor{color43}{rgb}{0.59,0.25,0.0}
\definecolor{color45}{rgb}{1.0,0.3,0}

\psline[linewidth=0.024cm,linecolor=color38](0.9,1.56)(0.9,-1.56)
\psline[linewidth=0.024cm,linecolor=color38](2.56,1.56)(2.56,-1.56)
\psline[linewidth=0.024cm,linecolor=color38](3.34,1.6)(4.12,1.6)
\psline[linewidth=0.024cm,linecolor=color38](4.14,1.56)(4.14,-1.56)
\psline[linewidth=0.024cm,linecolor=color38](3.36,-1.56)(4.14,-1.56)
\psline[linewidth=0.024cm,linecolor=color38](4.14,0.0)(3.34,0.0)
\psline[linewidth=0.024cm,linecolor=color38](8,1.56)(8,-1.56)
\psline[linewidth=0.024cm,linecolor=color38](9.6,1.56)(9.6,-1.56)
\psframe[linewidth=0.06,linecolor=color45,dimen=outer](3.3,1.6)(0.1,-1.6)
\psframe[linewidth=0.06,linecolor=color45,dimen=outer](10.4,1.6)(7.2,-1.6)
\psline[linewidth=0.06cm,linecolor=color45](8.8,1.6)(8.8,-1.6)
\psline[linewidth=0.06cm,linecolor=color45](7.2,0.0)(10.4,0.0)
\psline[linewidth=0.06cm,linecolor=color45](1.7,1.6)(1.7,-1.6)
\psline[linewidth=0.06cm,linecolor=color45](0.1,0.0)(3.3,0.0)

\psline[linewidth=0.024cm,linecolor=color38](7.96,-0.8)(10.36,-0.8)
\psline[linewidth=0.024cm,linecolor=color38](0.18,-0.78)(4.12,-0.78)
\psline[linewidth=0.024cm,linecolor=color38](0.18,0.82)(4.14,0.82)
\psellipse[linewidth=0.04,linecolor=color41,dimen=outer](7.25,0.03)(0.13,1.57)
\psline[linewidth=0.024cm,linecolor=color38,linestyle=dashed,dash=0.16cm 0.16cm](7.96,-0.8)(7.2,-0.8)
\psline[linewidth=0.024cm,linecolor=color38,linestyle=dashed,dash=0.16cm 0.16cm](7.96,0.8)(7.2,0.8)
\psline[linewidth=0.024cm,linecolor=color38,linestyle=dashed,dash=0.16cm 0.16cm](10.44,1.6)(11.14,1.6)
\psline[linewidth=0.024cm,linecolor=color38,linestyle=dashed,dash=0.16cm 0.16cm](11.16,1.58)(11.16,-1.6)
\psline[linewidth=0.024cm,linecolor=color38,linestyle=dashed,dash=0.16cm 0.16cm](10.44,-1.6)(11.16,-1.6)
\psline[linewidth=0.024cm,linecolor=color38,linestyle=dashed,dash=0.16cm 0.16cm](10.38,-0.8)(11.16,-0.8)
\psline[linewidth=0.024cm,linecolor=color38,linestyle=dashed,dash=0.16cm 0.16cm](10.46,-0.02)(11.14,-0.02)
\psline[linewidth=0.024cm,linecolor=color38,linestyle=dashed,dash=0.16cm 0.16cm](10.38,0.8)(11.14,0.8)
\psdots[dotsize=0.3,linecolor=color43](0.1,-1.6)
\psdots[dotsize=0.3,linecolor=color43](1.7,-1.6)
\psdots[dotsize=0.3,linecolor=color43](3.3,-1.6)
\psdots[dotsize=0.3,linecolor=color43](0.1,-0.0)
\psdots[dotsize=0.3,linecolor=color43](1.7,-0.0)
\psdots[dotsize=0.3,linecolor=color43](3.3,0.0)
\psdots[dotsize=0.3,linecolor=color43](3.3,1.6)
\psdots[dotsize=0.3,linecolor=color43](1.7,1.6)
\psdots[dotsize=0.3,linecolor=color43](0.1,1.6)
\psdots[dotsize=0.3,linecolor=color43](7.2,-1.6)
\psdots[dotsize=0.3,linecolor=color43](8.8,-1.6)
\psdots[dotsize=0.3,linecolor=color43](10.4,-1.6)
\psdots[dotsize=0.3,linecolor=color43](7.2,-0.0)
\psdots[dotsize=0.3,linecolor=color43](8.8,-0.0)
\psdots[dotsize=0.3,linecolor=color43](10.4,-0.0)
\psdots[dotsize=0.3,linecolor=color43](10.4,1.6)
\psdots[dotsize=0.3,linecolor=color43](8.8,1.6)
\psdots[dotsize=0.3,linecolor=color43](7.2,1.6)
\psline[linewidth=0.024cm,linecolor=color38](8.0,0.8)(10.3,0.8)
\pscircle[linewidth=0.024,linecolor=color43,dimen=outer](0.5,-1.2){0.56}
\pscircle[linewidth=0.024,linecolor=color43,dimen=outer](2.1,-1.2){0.56}
\pscircle[linewidth=0.024,linecolor=color43,dimen=outer](3.7,-1.2){0.56}
\pscircle[linewidth=0.024,linecolor=color43,dimen=outer](7.6,-1.2){0.56}
\pscircle[linewidth=0.024,linecolor=color43,dimen=outer](9.2,-1.2){0.56}
\pscircle[linewidth=0.024,linecolor=color43,dimen=outer](10.8,-1.2){0.56}
\psline[linewidth=0.05cm,linecolor=color41,arrowsize=0.05291667cm 2.0,arrowlength=1.4,arrowinset=0.4]{<-}(4.14,2.0)(4.14,2.44)
\psline[linewidth=0.05cm,linecolor=color41,arrowsize=0.05291667cm 2.0,arrowlength=1.4,arrowinset=0.4]{<-}(0.16,2.0)(0.16,2.44)
\psline[linewidth=0.05cm,linecolor=color41](1.34,2.44)(0.16,2.44)
\psline[linewidth=0.05cm,linecolor=color41](4.14,2.44)(2.94,2.44)
\psline[linewidth=0.05cm,linecolor=color41,arrowsize=0.05291667cm 2.0,arrowlength=1.4,arrowinset=0.4]{<-}(10.4,2.0)(10.4,2.44)
\psline[linewidth=0.05cm,linecolor=color41](10.4,2.44)(9.9,2.44)
\psline[linewidth=0.05cm,linecolor=color41,arrowsize=0.05291667cm 2.0,arrowlength=1.4,arrowinset=0.4]{<-}(7.98,2.0)(7.98,2.44)
\psline[linewidth=0.05cm,linecolor=color41](8.5,2.44)(7.98,2.44)
\psline[linewidth=0.05cm,linecolor=color41,arrowsize=0.05291667cm 2.0,arrowlength=1.4,arrowinset=0.4]{<-}(7.18,-1.98)(7.18,-2.46)
\psline[linewidth=0.05cm,linecolor=color41,arrowsize=0.05291667cm 2.0,arrowlength=1.4,arrowinset=0.4]{<-}(11.16,-1.96)(11.16,-2.44)
\psline[linewidth=0.05cm,linecolor=color41](8.36,-2.44)(7.16,-2.44)
\psline[linewidth=0.05cm,linecolor=color41](11.18,-2.44)(9.98,-2.44)
\psellipse[linewidth=0.04,linecolor=color41,dimen=outer](4.15,0.05)(0.13,1.57)
\psline[linewidth=0.04cm,linecolor=color41,arrowsize=0.05291667cm 2.0,arrowlength=1.4,arrowinset=0.4]{<-}(4.6,-0.0)(5.1,0.0)
\psline[linewidth=0.04cm,linecolor=color41,arrowsize=0.05291667cm 2.0,arrowlength=1.4,arrowinset=0.4]{<-}(6.8,-0.0)(6.3,-0.0)

\put(8.72,2.37){\tiny{Dirichlet}}
\put(1.65,2.37){\tiny{Dirichlet}}
\put(8.35,-2.35){\tiny{Non-dynamical}}
\put(8.85,-2.65){\tiny{fields}}
\put(5.25,-0.1){\tiny{Periodic}}
\put(0.1,1.8){\tiny{0}}
\put(0.85,1.8){\tiny{1}}
\put(1.65,1.8){\tiny{2}}
\put(2.5,1.8){\tiny{3}}
\put(3.35,1.8){\tiny{4}}
\put(4.1,1.8){\tiny{5}}
\put(-0.3,1.9){\tiny{$\frac{x_0}{a}$}}
\put(4.4,1.7){\tiny{$ = \frac{T}{a}$}}
\put(7.5,1.9){\tiny{$  \frac{x_0}{a}$}}
\put(7.95,1.8){\tiny{0}}
\put(8.75,1.8){\tiny{1}}
\put(9.55,1.8){\tiny{2}}
\put(10.35,1.8){\tiny{3}}
\put(10.55,1.8){\tiny{$=\frac{T}{a}$}}
\put(4.5,1.3){\tiny{$\mathbf{\frac{x}{a}}$}}
\put(4.3,1.5){\tiny{\bf 0}}
\put(4.3,0.7){\tiny{\bf 1}}
\put(4.3,-0.1){\tiny{\bf 2}}
\put(4.3,-0.9){\tiny{\bf 3}}
\put(4.3,-1.7){\tiny{\bf 4}}
\put(4.5,-1.7){\tiny{$\mathbf{=\frac{L}{a}} $}}
\put(11.5,1.3){\tiny{$\mathbf{\frac{x}{a}}$}}
\put(11.3,1.5){\tiny{\bf 0}}
\put(11.3,0.7){\tiny{\bf 1}}
\put(11.3,-0.1){\tiny{\bf 2}}
\put(11.3,-0.9){\tiny{\bf 3}}
\put(11.3,-1.7){\tiny{\bf 4}}
\put(11.5,-1.7){\tiny{$\mathbf{=\frac{L}{a}} $}}

\put(6.8, -2.){\tiny{$\frac{y_0}{a}$}}
\put(7.2, -2.){\tiny{0}}
\put(8.75, -2.){\tiny{1}}
\put(10.35, -2.){\tiny{2}}
\put(-0.3, -2.05){\tiny{$\frac{y_0}{a}$}}
\put(0.1, -2.){\tiny{0}}
\put(1.65, -2.){\tiny{1}}
\put(3.25, -2.){\tiny{2}}
\put(-0.6,1.5){\tiny{$\mathbf{\frac{y}{a}}$}}
\put(-0.3,1.5){\tiny{\bf 0}}
\put(-0.3,-0.1){\tiny{\bf 1}}
\put(-0.3,-1.7){\tiny{\bf 2}}
\put(6.55,1.5){\tiny{$\mathbf{\frac{y}{a}}$}}
\put(6.85,1.5){\tiny{\bf 0}}
\put(6.85,-0.1){\tiny{\bf 1}}
\put(6.85,-1.7){\tiny{\bf 2}}

\end{pspicture} 
}
\end{center}
\caption{\label{fig1} 
The figure shows a 2-dimensional section of the SF with staggered fermions with $L/a=4$ and $T'/L=1$. 
The left panel shows the case $T'= T-a$ ($s=-1$) the right panel corresponds to $T'= T+a$ ($s=1$).
Thin lines represent the lattice on which the one-component staggered fermions live. 
The Dirac spinors are reconstructed from those one-component fields 
residing inside the circles and live on the effective lattice depicted by the thick lines.}

\end{center}
\end{figure}
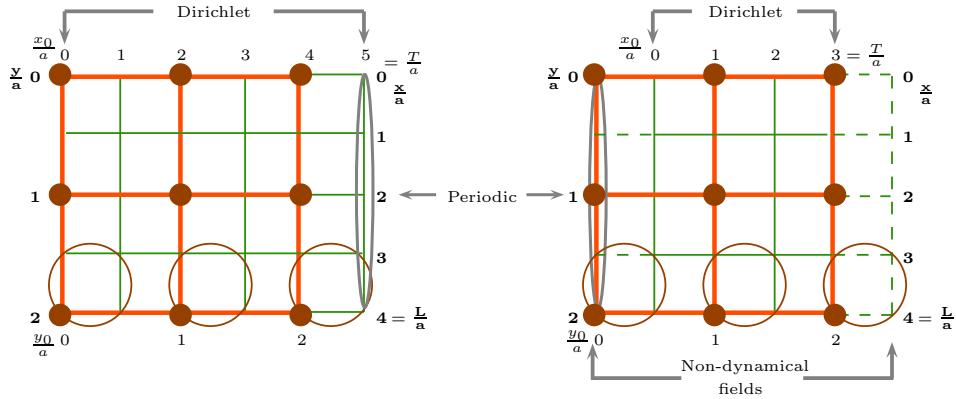

A peculiarity in the formulation of the Schr\"odinger functional with staggered quarks has 
first been noticed in~\cite{Miyazaki:1994nu} and can also be traced back to the
properties of the transfer matrix~\cite{Sharatchandra:1981si}: the lattice sizes needed in this case have even 
spatial extent $L/a$ and odd temporal extent $T/a$. To understand this constraint, 
one needs to recall that the 4 flavours of Dirac (i.e.~four-component) spinors are reconstructed
from the staggered one-component fields living on the $2^4=16$ corners of the elementary 
hypercubes of the lattice. The reconstructed fields may be imagined to live on a coarse 
lattice with the doubled lattice spacing. The constraint arises from having 
to construct an integer multiple of four Dirac spinors from those staggered one-component 
fields which are integration variables in the functional integral (cf. the left panel of~figure~\ref{fig1}  
for an illustration in 2 dimensions).
Equivalently, the lattice geometry must be such that an integer number of hypercubes 
with dynamical field components is obtained.  Note that the Dirichlet conditions  
for the one-component fields at $x_0=0$ and $x_0=T$ naturally lead to a Dirichlet condition 
for half of the Dirac spinors living at the boundaries. 
This implicitly defines projectors in spin-flavour space, the form of which depends  
on the details of the reconstruction of the Dirac spinors. 
However, there is enough freedom to obtain four flavours of quarks 
satisfying the standard boundary conditions in terms of the projectors 
$P_\pm=\frac12(1\pm\gamma_0)$~\cite{Sint:notes1994,Heller:1997pn}.
The reconstructed four-spinors thus live on a coarser lattice with size $(L/2a)^3\times(T/a - 1)/2$.

A slightly less intuitive option to reconstruct the four-component spinors
is  illustrated in the right panel of figure~\ref{fig1}. It corresponds to an extension of the fine lattice by 
a time slice beyond each of the time boundaries. The part of the four-component spinors living on these added time-slices 
correspond to the non-Dirichlet components at the boundaries, which need not be
dynamical field components\footnote{This is in fact the situation in the SF with Wilson quarks,
where the non-Dirichlet components at the boundaries are completely decoupled from the dynamical 
field components~\cite{Sint:1993un}.}.
To treat this second case we also consider the coarser lattice with size $(L/2a)^3\times(T/a + 1)/2$,
and, combining both options, we set
\begin{equation}
  T'=T+sa, \qquad s=\pm 1.
\end{equation}
Note that in the absence of staggered quarks we may also set $s=0$ to recover the standard framework
for the pure gauge theory. In the following we will thus try to take the continuum limit 
at fixed ratio, $\rho=T'/L$. In particular, for the computation of the SF coupling we set $\rho=1$ exactly. 
This redefines the approach to the continuum limit, and we are thus led to discuss on-shell O($a$) improvement 
for this modified set-up.

\subsection{The equations of motion and the background field}

The background field is the field configuration which minimises the lattice action. Since the
latter depends on $\ct$, any change of this coefficient may modify the minimal action configuration.
The equations of motion can easily be generalised to this case by simply including the 
weight factors $w_{\mu\nu}(x)\equiv w[P_{\mu\nu}(x)]$ in the definition of the 
covariant divergence, Eq.~(\ref{eq:covdiv}),
\begin{eqnarray}
  d_w^\ast P(x,\mu) &=& \sum_{\nu=0}^3\Bigl\{w_{\mu\nu}(x) P_{\mu\nu}(x)  \nonumber\\
     && - w_{\mu\nu}(x-a\hat\nu) U_\nu^\dagger(x-a\hat\nu)P_{\mu\nu}(x-a\hat\nu)U_\nu^{}(x-a\hat\nu)\Bigr\}.
\end{eqnarray}
The lattice action is stationary if and only if the traceless antihermitian part of 
$d_w^*P$ vanishes, i.e.~the equations of motion read
\begin{equation}
   d_w^\ast P(x,\mu) - d_w^\ast P(x,\mu)^\dagger 
   -\frac{1}{N}\tr\left\{ d_w^\ast P(x,\mu) - d_w^\ast P(x,\mu)^\dagger \right\} =0.
\label{eq:eomw}
\end{equation}
In order to solve the equations we choose the temporal gauge, setting all temporal links to unity, $U_0(x) = 1$.  
Since the modification of the standard framework with $T'=T$ amounts to an O($a$) effect, 
it seems plausible that the solution of the equations of motion will
again be Abelian and spatially constant, at least for large enough lattice sizes.  Hence, we make the ansatz
\begin{equation}
  V(x,k)= \exp[aB_k(x_0)],
\end{equation}
and try to solve Eq.~(\ref{eq:eom}) for a colour diagonal and spatially 
constant field $B_k(x_0)$, subject to the boundary conditions
\begin{equation}
   B_k(0)=C^{}_k= i\phi/L, \qquad B_k(T)=C'_k=i\phi'/L,
\end{equation}
The coefficient $\ct$ is left as a free parameter, 
which will be fixed later by demanding that the action is O($a$) improved.
The field tensor associated with the plaquette field is defined by
\begin{equation}
   P_{\mu\nu}(x)\vert_{U\rightarrow V} = \exp[a^2G_{\mu\nu}(x_0)],
\end{equation}
and is thus related to $B_k(x_0)$ by the forward lattice derivative,
\begin{equation}
  G_{0k}(x_0) = \partial_0 B_k(x_0) = \frac{1}{a}\left[B_k(x_0 + a) - B_k(x_0)\right],
  \label{eq:G0kdef}
\end{equation}
with all other components being zero. 
The boundary conditions are equal for all spatial indices $k=1,2,3$, 
so that we introduce the notation
\begin{equation}
  f(x_0) = G_{0k}(x_0).
\end{equation}
Note that $f(x_0)$ is, for fixed $x_0$, an anti-hermitian diagonal matrix in colour space with trace
zero, i.e.~its colour components $f_\alpha(x_0)$ are purely imaginary and must sum to zero. 
Eqs.~(\ref{eq:eomw}) now reduce to $T/a$ equations of the form,
\begin{equation}
   M(x_0) - \frac{1}{N}\tr\left[M(x_0)\right]=0, \qquad  a\le x_0\le T-a,
\label{eq:eom_m}
\end{equation}
where the trace is over colour and the diagonal $N\times N$-matrices $M(x_0)$ are given by
\begin{equation}
M(x_0) = \left\{\begin{array}{ll} 
         \sinh \left[a^2f(a)\right] - \ct \sinh \left[a^2f(0)\right]    
	  & \text{if } x_0= a, \\
          \ct \sinh\left[a^2f(T - a)\right] - \sinh\left[a^2f(T - 2a)\right] 
          & \text{if } x_0=T-a,\\ 
           \sinh\left[a^2f(x_0)\right]- \sinh\left[a^2f(x_0 - a)\right]
	  & \text{if } a < x_0 < T-a. 
             \end{array}
        \right.
\end{equation}
Expanding $f(x_0)$ in a power series in $a$,
\begin{equation}
   f(x_0) = \sum_{n = 0}^\infty \left(\frac{a}{L}\right)^n f^{(n)}(x_0),
  \label{eq:f_expansion}
\end{equation}
and inserting this expansion in the equations of motion (\ref{eq:eom_m}), 
one may show by induction that, for all $n$, 
\begin{equation}
  f^{(n)}(x_0) = \left\{\begin{array}{ll}
               f^{(n)} +\Delta f^{(n)} 
               &  \text{if  $x_0 = 0, T -a$}\\ 
               f^{(n)}   
                &  \text{if $ a \le x_0 < T -a$}. 
                 \end{array}
                  \right.
\end{equation}
Hence one concludes that $f(x_0)$ is equal to the $x_0$-independent constant matrix $f$ 
except for a jump at the time boundaries by $\Delta f$.
To proceed further one needs to go back to the background field $B_k(x_0)$ in the temporal gauge.
One finds,
\begin{eqnarray}
 B_k(0)   &=& C_k \nonumber \\
 B_k(x_0) &=& \left(x_0  - \frac T2\right) f + \frac{C_k + C_k'}{2} 
               \qquad \text{if $a \le x_0 \le T -a$} \label{eq:BFmod}\\
  B_k(T)  &=& C_k',\nonumber
\end{eqnarray}
and the jump in the field tensor at the boundaries is given by
\begin{equation}
  \Delta f = \frac{C'_k-C_k^{}}{2a}-\frac{T}{2a} f.
\end{equation}
Thus, for given boundary gauge fields and lattice geometry the background field and its field tensor are 
determined by the traceless and diagonal colour matrix $f$, which still needs to be computed.

We now specialise to $N=3$ colours and insert the matrices $f$ and $\Delta f$
in Eq.~(\ref{eq:eom_m}). While $M(x_0)=0$ for all $a<x_0<T-a$, the 2 equations 
for $M(a)$ and $M(T-a)$ are identical and read, in terms of the colour components
$f_\alpha$,
\begin{eqnarray}
0 &=& \frac{1}{3} \sum_{\beta = 1}^3 \left\{  
    \ct \sinh\left \{ a^2f_\beta \left\lbrack 1 + \frac s2 
    - \frac{T'}{2a}\right\rbrack  + ia \frac{\phi'_\beta - \phi_\beta^{}}{2L}\right\} 
    - \sinh\left\{a^2 f_\beta  \right\} \right\}   
    \nonumber \\
&&    -\ct \sinh\left \{ a^2f_ \alpha  \left\lbrack 1 + \frac s2  
   - \frac{T'}{2a}\right\rbrack  + ia \frac{\phi_\alpha' - \phi_\alpha^{}}{2L}\right\} 
   + \sinh\left\{a^2 f_\alpha  \right\} ,
\end{eqnarray}
for $\alpha = 1,2,3$. Taking into account that 
$\phi'_1- \phi_1^{} = -2(\phi_2' - \phi_2^{}) = -2(\phi_3' - \phi_3^{})$ [cf.~Eqs.~(\ref{eq:phi_phiprime})]
one concludes that $f_2$ and $f_3$ satisfy the same equation and hence $f_2 = f_3$. Tracelessness
of $f$ then implies 
\begin{equation}
   f = f_2\times\diag(-2,1,1),
 \label{eq:f_from_f2}
\end{equation}
so that we are left with a single equation for $f_2$. Introducing the real 
dimensionless variable $\varphi$ through
\begin{equation}
   f_2=i\varphi/L^2,
 \label{eq:varphi_def}
\end{equation}
the equation to be solved for $\varphi$ is
\begin{equation}
 \ct \sin\left\{2\frac{a^2}{L^2}K(\varphi)\right\}+  
 \ct \sin\left\{\frac{a^2}{L^2}K(\varphi)\right\} 
 -\sin\left\{2\frac{a^2}{L^2}\varphi\right\}-  
  \sin\left\{\frac{a^2}{L^2}\varphi\right\} = 0,
\label{eq:varphi}
\end{equation}
with
\begin{equation}
   K(\varphi) = \varphi\left[1+\frac{s}2-\frac{\rho L}{2a}\right]+\frac{L}{2a}\left(\phi'_2-\phi_2^{}\right),\qquad
   \rho= T'/L.
\end{equation}
Next we expand $\varphi$ in powers of $a/L$ in analogy to Eq.~(\ref{eq:f_expansion})
and solve Eqs.~(\ref{eq:varphi}) order by order in $a/L$. 
For the first 3 coefficients, we find,
\begin{eqnarray}
   \varphi^{(0)} &=& \rho^{-1} \left(\phi'_2-\phi_2^{}\right) =   \rho^{-1} \left(\eta + \frac{\pi}{3}\right)
   \label{eq:varphi_0}\\[1ex]
   \varphi^{(1)} &=& \varphi^{(0)} k_s ,\qquad k_s = \frac{(2+s)\ct-2}{\rho \ct}, 
   \label{eq:ct0dep}\\
   \varphi^{(2)} &=& \varphi^{(0)} k_s^2.
\end{eqnarray}
Without further assumptions on $\ct$ or $s$ there are corrections
to all orders in $a/L$. We note, however, that for the standard
case with $s=0$ and $\ct=1$, Eq.~(\ref{eq:varphi}) 
reduces to $K(\varphi)=\varphi$, which is exactly solved by $\varphi=\varphi^{(0)}$.
In the general case one may obtain precise approximate solutions 
by pushing the Taylor expansion to higher orders. Alternatively, one may, 
for given numerical values of $s$, $L/a$, $\rho$, $\ct$ and $\eta$, 
find a numerical solution for $\varphi$ (and thus for $f$) by applying the Newton procedure 
to Eq.~(\ref{eq:varphi}). 

\subsection{Choice of $\ct$}

The free parameter $\ct$ of the background field can now be determined by 
demanding that the classical action is O($a$) improved. This can be done analytically, as
the dependence of the background field on $\ct$ is known to O($a$) 
from Eq.~(\ref{eq:ct0dep}). First we insert the background field (\ref{eq:BFmod}) into the action, 
and set $T=T'-sa$,
\begin{eqnarray}
  S[V] &=& \frac{-3[T'-(2+s)a]L^3}{g_0^2}
         \sum_{\alpha=1}^3\left\{\frac{2}{a^2}\sinh\left[\frac{a^2}{2}f_\alpha \right] \right\}^2 
  \nonumber\\
   && \hbox{} - \frac{6a L^3 \ct}{g_0^2}
        \sum_{\alpha=1}^3
        \left\{\frac{2}{a^2}\sinh\left[\frac{a^2}{2}(f_\alpha +\Delta f_\alpha)\right] \right\}^2. 
\label{eq:ct_dependence}
\end{eqnarray}
Then, using the relations~(\ref{eq:f_from_f2},\ref{eq:varphi_def}), and
\begin{equation}
   f+\Delta f = \frac{i}{L^2} K(\varphi) \times\diag(-2,1,1),
\end{equation}
the action can be written as a function of $\varphi$, viz.
\begin{eqnarray}
 S[V] &=& -\frac{24 L^3}{g_0^2 a^3} \bigg\{\left[1+\frac{s}2-\frac{\rho L}{2a}\right]
           \left[\sin^2\left(\frac{a^2}{L^2}\varphi\right)+2\sin^2\left(\frac{a^2}{2L^2}\varphi\right)\right] \nonumber\\
&& \hphantom{012345} - \ct \left[\sin^2\left(\frac{a^2}{L^2}K(\varphi)\right)+2\sin^2\left(\frac{a^2}{2L^2}K(\varphi)\right)\right]
   \bigg\} 
\end{eqnarray}
Then, with
\begin{equation}
  \varphi=\varphi^{(0)}\left[1+ \frac{a}{L} k_s + \frac{a^2}{L^2} k_s^2 +\rmO(a^3)\right],
\qquad
  K(\varphi) = \frac{\varphi^{(0)}}{\ct}\left[1+ \frac{a}{L} k_s   +\rmO(a^2)\right],
\end{equation}
and Taylor expanding in $a/L$, one obtains
\begin{equation}
  S[V] = \frac{18 \rho}{g_0^2}\left(\varphi^{(0)}\right)^2\left\{1 + \frac{a}{L}k_s +\rmO(a^2)\right\}
       = S_{\rm cont}[B]\left\{1 + \frac{a}{L}k_s +\rmO(a^2)\right\}.
\end{equation}
In order to reduce the corrections to O($a^2$) the coefficient $k_s$, as defined in Eq.~(\ref{eq:ct0dep}) 
must vanish, which is achieved by setting
\begin{equation}
   \ct = \frac{2}{2+s} = \left\{\begin{array}{ll} 
	   \frac23 & \text{if } s=1, \\
           1 & \text{if } s=0,\\ 
           2 & \text{if } s=-1.  
             \end{array}
        \right.
\label{eq:ct0}
\end{equation}
Remarkably, with this choice for $\ct$, the action of the background field is
not just O($a$) improved but cutoff effects only start at O($a^4$), 
in all three cases $s=\pm 1,0$. In fact, the background field is only mildly
distorted by lattice artefacts, as shown by the Taylor expansion,
\begin{equation}
 \left.\varphi\right\vert_{\ct=2/(2+s)} = \varphi^{(0)} + \left(\frac{a}{L} \right)^5 \varphi^{(5)} +
\left(\frac{a}{L}\right)^9\varphi^{(9)} +
\left(\frac{a}{L}\right)^{10}\varphi^{(10)}  +{\rm O}\left( a^{13}\right),
\label{eq:varphi_aexpansion}
\end{equation}
where $\varphi^{(0)}$ is given in Eq.~(\ref{eq:varphi_0}) and
\begin{eqnarray}
\varphi^{(5)} &=& -\frac{(\phi_2' - \phi_2)^3}{8\rho^4}s(s+4)(s+2), 
   \label{eq:varphi5}\\
\varphi^{(9)} &=& -\frac{(\phi_2' - \phi_2)^5}{120\rho^6}s(s+4)(s+2)\left(\tfrac{79}{16}s^2 +\tfrac{79}{4}s +17 \right ),
   \label{eq:varphi9}\\
\varphi^{(10)} &=& \frac{3(\phi_2' - \phi_2)^5}{2^6\rho^7}s^2(s+4)^2(s+2)^2 = \frac{3}{\varphi^{(0)}}\left(\varphi^{(5)}\right)^2.
   \label{eq:varphi10}
\end{eqnarray}
The expansion to this order yields rather precise approximations to the full solution of Eq.~(\ref{eq:varphi}),
For instance, with $\rho=1$, we conclude that $\varphi/\varphi^{(0)}$ 
is obtained with double precision provided $L/a > 12$ and $L/a>21$ 
for $s=-1$ and $s=1$ respectively. 

In Section~4 we will also need the $\eta$-derivative of the background field, 
i.e.~of $\varphi$. Given $\varphi$, this can be obtained by differentiating 
Eq.~(\ref{eq:varphi}) with respect to $\eta$. 
Alternatively one may obtain the $a/L$-expansion by differentiating 
the coefficients in Eqs.~(\ref{eq:varphi5}-\ref{eq:varphi10}). Recalling that
$\phi_2'-\phi_2^{}=\eta+\pi/3$, the $\eta$-derivatives of the coefficients are given by
\begin{equation}
\partial_\eta\varphi^{(5)} = \frac{3\varphi^{(5)}}{\phi_2' - \phi_2},\qquad
\partial_\eta\varphi^{(9)} = \frac{5\varphi^{(9)}}{\phi_2' - \phi_2},\qquad
\partial_\eta\varphi^{(10)} = \frac{5\varphi^{(10)}}{\phi_2' - \phi_2}.
\end{equation}

\subsection{A numerical check}

The above solution of the classical field equations was obtained using the 
hypothesis that the background field is Abelian and spatially constant.
While this can be rigorously established in the standard set-up, we here
rely on the plausibility of the assumption that an O($a$) change
in the set-up can only have a small impact, at least for large enough lattice sizes.
To check the hypothesis on small size lattices ($L/a =4-8$), we have tried to find
the absolute minimum of the action numerically, by starting from a random
gauge configuration. Sweeping through the lattice we minimised the plaquettes 
with respect to the SU(3) link variables. In order to avoid getting trapped in
secondary minima of the action we also used over-relaxation steps which change the 
configuration whilst leaving the action invariant. The obtained results agreed numerically 
very well with the expectation from the Abelian background field. Furthermore, the background
field itself was found to be of the expected form after transformation 
to the temporal gauge. We take this as evidence that the true minimal action configuration
is indeed the Abelian and spatially constant solution described above.

  \section{One-loop results}

The perturbation expansion is analogous to the calculation in SU(2) which has been
described in the original paper~\cite{Luscher:1992an} and we will be rather brief on 
the technical details. Given the one-loop results we obtain the
boundary improvement coefficients to this order and we can check remaining
lattice artefacts in the step-scaling function for the coupling and in $\bar v(L)$.

\subsection{The running coupling at one-loop order}

So far our considerations have been purely classical. Dealing with quantum corrections the
improvement coefficient $\ct$ will become a function of the bare coupling, with an expansion
of the form,
\begin{equation}
   \ct(g_0)=\ct^{(0)}+g_0^2\ct^{(1)}+\rmO(g_0^4).
\end{equation}
Our classical considerations in the previous section correspond to setting $g_0=0$ i.e.~$\ct=\ct^{(0)}$,
and $\ct^{(0)}$ is thus given by Eq.~(\ref{eq:ct0}).
The perturbative expansion of the effective action on the lattice takes
the same form as in the continuum, Eq.~(\ref{eq:coup_exp}), with the first 2 terms given by
\begin{eqnarray}
\Gamma_0[B] &=& g_0^2 S[V]\vert_{\ct=\ct^{(0)}}\\
\Gamma_1[B] &=& \ct^{(1)}\Gamma_{0;\ct}[B] - \ln\det \Delta_0 +\frac12 \ln \det \Delta_1.
\end{eqnarray}
Here, the fluctuation operators $\Delta_0$ and $\Delta_1$ appear in the Gaussian parts
of the action in the Faddeev-Popov ghost and the gluon fields, respectively.
The counterterm $\propto \ct^{(1)}$ is specified by
\begin{equation}
  \Gamma_{0;\ct}[B] = \frac{\partial}{\partial\ct} \Gamma_0[B]\vert_{\ct=\ct^{(0)}} = 24\frac{L^3}{a^3} 
  \left[\sin^2\left(\frac{a^2}{L^2}K(\varphi)\right)+2\sin^2\left(\frac{a^2}{2L^2}K(\varphi)\right)\right],
\end{equation}
where $\varphi$ is the solution of Eq.~(\ref{eq:varphi}) with $\ct=\ct^{(0)}$.
Setting $T'=L$ (i.e.~$\rho=1$) the definition of the SF coupling on the lattice
\begin{equation}
  \bar{g}^2 (L) = \left.\frac{\partial_\eta\Gamma_0[B]}{\partial_\eta\Gamma[B]}\right\vert_{\eta=\nu=0},
  \label{eq:gbar_latdef}
\end{equation}
is such that $\bar{g}(L)=g_0$ holds exactly at lowest order. Hence, the normalisation constant,
\begin{equation}
  \partial_\eta \Gamma_0[B] = 12 \frac{L^2}{a^2}
  \left[
  \sin\left(\frac{a^2}{L^2}\varphi\right) + 2\sin\left(2\frac{a^2}{L^2}\varphi\right)  
  \right],
  \label{eq:normalisation}
\end{equation}
must be computed including the lattice artefacts and differs from its value in the continuum 
limit, $k=12\pi$ [cf.~Eq.~(\ref{eq:coupling})], by terms of O($a^4$).

To one-loop order the perturbative relation to the bare coupling then reads,
\begin{equation}
   \bar g^2(L) = g_0^2 + \left(m_1(L/a)  
   -\ct^{(1)}\left.\frac{\partial_\eta \Gamma_{0;\ct}[B]}{\partial_\eta\Gamma_0[B]}\right\vert_{\eta=0}\right) g_0^4 + \rmO(g_0^6),
\label{eq:gbar_pert}
\end{equation}
where the coefficient $m_1(L/a)$ is a sum of two contributions,
\begin{equation}
   m_1 = \left[h_0 - \mbox{$\frac 12$}h_1\right]_{\nu=0}.
  \label{eq:m1}
\end{equation}
which derive from the Fadeev-Popov ghost and gluon fluctuation operators,
\begin{equation}
   h_j = \left.\frac{\partial_\eta \ln\det \Delta_j}{\partial_\eta\Gamma_0[B]}\right\vert_{\eta=0}, \qquad j = 0,1.
  \label{eq:hj}
\end{equation}
We have computed these contributions numerically for lattice sizes ranging from $L/a=4$ to $L/a=64$, 
using two completely different techniques.  One of us has performed the computation 
along the lines of refs.~\cite{Luscher:1992an,PWeisz}, using recursion relations for the 
finite difference operators $\Delta_0$ and $\Delta_1$, in order to compute their determinants  
in each sector of fixed spatial momentum and colour quantum numbers.
In addition, an independent computation was carried out based on the automated perturbation theory described 
in refs.~\cite{Hart:2004bd,Takeda:2008rr}. Perfect numerical agreement was found between the two methods, up to rounding errors.
We also verified gauge invariance of $m_1(L/a)$ by computing it at different values of the gauge fixing parameter. 
For future reference the numerical results for $m_1(L/a)$ are collected in Tables~\ref{m1_data_even} and~\ref{m1_data_odd} of Appendix~A, 
for all three choices of $s=\pm 1,0.$. To estimate the numerical precision we have compared the 
two calculations and find agreement for 15 digits at $L/a=4$ which reduces to about 12 digits at $L/a=64$, due to the accumulation
of rounding errors.

\subsection{Determination of $\ct^{(1)}$}

From Symanzik's analysis \cite{Symanzik:1979ph,Symanzik:1982dy} 
of the cutoff dependence of Feynman diagrams on the lattice, 
one expects $m_1(L/a)$ to have an asymptotic expansion of the form
\begin{equation}
  m_1(L/a)\stackrel{L/a\to \infty}{\sim}\sum_{n=0}^\infty (a/L)^n
  \left[r_n +s_n\ln(L/a)\right].
\label{eq:as-ex}
\end{equation}
The series is logarithmically divergent, since $m_1(L/a)$ relates a renormalised to 
a bare coupling. For the coefficient of the divergence one expects
$s_0=2b_0$, where $b_0$ is the one-loop coefficient of the SU(3) $\beta$-function, Eq.~(\ref{eq:SU3beta}).
The divergence could be eliminated by renormalising the bare coupling $g_0$ 
e.g.~by minimal subtraction of logarithms~\cite{Collins,Sint:1995ch}.
If one passes directly to the $\overline{\rm MS}$-scheme of dimensional regularisation,
one finds that the coefficient $c_1$ in Eq.~(\ref{eq:c1}) is related to $r_0$ in Eq.~(\ref{eq:as-ex})
through
\begin{eqnarray}
 c_1 &=&  d_1(1)\vert_{N=3} - 4\pi r_0, \\
d_1(1) &=& -\frac{\pi}{2N} + k_1N, \qquad k_1 = 2.135730074078457(2), \nonumber
\end{eqnarray}
where the coefficient $d_1(1)$ is taken from ref.~\cite{Luscher:1995nr}. Within the numerical precision
we reproduce the known result for $r_0$ of ref.~\cite{Luscher:1993gh} in all three cases, $s=\pm 1,0$.

The coefficient $r_1$ multiplies cutoff effects of order $a/L$ which ought to be cancelled by
correctly choosing $\ct^{(1)}$. Hence we need to evaluate the coefficient of
$\ct^{(1)}$ in Eq.~(\ref{eq:gbar_pert}) to O($a$). To this order, the $\ct$-dependence of $\Gamma_0$ 
can be inferred from Eq.~(\ref{eq:ct_dependence}) and is given by
\begin{equation}
   \Gamma_{0;{\ct}}[B] = \partial_{\ct} \left[g_0^2 S[V]\right]_{\ct=\ct^{(0)}}  
    = g_0^2 S_{\rm cont}[B]\frac{a}{L} \left[\partial_{\ct} k_s\right]_{\ct^{}=\ct^{(0)}} +\rmO(a^2),
 \label{eq:Gam_ct_as}
\end{equation}
and $\partial_{\ct} k_s= 2/(\rho\ct^2)$.  This shows explicitly 
that this term is of O($a$). Performing the $\eta$-derivative
and requiring the absence of O($a$) terms in the one-loop relation between the couplings, 
leads to the condition,
\begin{equation}
   \ct^{(1)} = \rho \frac{r_1}{2}\left(\ct^{(0)}\right)^2. 
   \label{eq:ct1_formula}
\end{equation}
Note that the coefficient $s_1$ of the term $(a/L)\times \ln(L/a)$ is expected to
vanish, provided tree-level improvement is correctly implemented.
Following the method presented in the appendix of \cite{Bode:1999sm}, the 
first few coefficients were extracted numerically. 
The outcome of this analysis for the cases $s=0,\pm 1$ is shown in Table~\ref{table_r0s0r1s1}.
In this table, coefficients shown with no errors have been assumed, entries $\times$ have also been 
fitted but are not listed here and the terms $\sim$ are included one at a time for 
the error analysis.
\begin{table}[ht]
\begin{center}
\begin{tabular}{|c|l|l|l|l|l|l|l|l|}
\hline
&$s_0$&$r_0$&$s_1$&$r_1$&$s_2$&$r_2$&$s_3$&$r_3$\\\cline{1-9}
\hline
\hline
&$0.13931(4)$&$0.3683(3)$&$-0.001(15)$&$-0.23(7)$&
$\times$&$\times$&$\sim$&$\sim$\\\cline{2-9}
$s=-1$&$22/(4\pi)^2$&$0.368283(3)$&$-0.0004(10)$&$-0.230(6)$&
$\times$&$\times$&$\sim$&$\sim$\\\cline{2-9}
&$22/(4\pi)^2$&$0.3682818(7)$&$0$&$-0.2318(3)$&
$\times$&$\times$&$\sim$&$\sim$\\\cline{2-9}
\hline
\hline
&$0.13931(3)$&$0.3683(2)$&$-0.001(13)$&$-0.17(7)$&
$\times$&$\times$&$\sim$&$\sim$\\\cline{2-9}
$s=0$&$22/(4\pi)^2$&$0.368283(2)$&$-0.0003(9)$&$-0.176(6)$&
$\times$&$\times$&$\sim$&$\sim$\\\cline{2-9}
&$22/(4\pi)^2$&$0.3682817(7)$&$0$&$-0.1779(3)$&
$\times$&$\times$&$\sim$&$\sim$\\\cline{2-9}
\hline
\hline
&$0.13931(2)$&$0.3683(1)$&$0.001(7)$&$0.12(3)$&
$\times$&$\times$&$\sim$&$\sim$\\\cline{2-9}
$s = 1$&$22/(4\pi)^2$&$0.368280(1)$&$0.0007(5)$&$0.120(3)$&
$\times$&$\times$&$\sim$&$\sim$\\\cline{2-9}
&$22/(4\pi)^2$&$0.368283(1)$&$0$&$0.1232(4)$&
$\times$&$\times$&$\sim$&$\sim$\\\cline{2-9}
\hline

\end{tabular}

\caption{\label{table_r0s0r1s1} Asymptotic expansion coefficients in  $m_1(L/a)$, for $s=\pm1,0$.}
\end{center}
\end{table} 
As can be seen from the results, all expectations are confirmed within the errors.
In particular we confirm the known result for the case $s=0$~\cite{Luscher:1993gh}, 
and the universal coefficients $r_0$ and $s_0$ agree for all choices of the parameter $s$. 
As expected, $s_1$ is found to be compatible with zero. 
Applying formula~(\ref{eq:ct1_formula}) with $\rho=1$ we obtain
\begin{equation}
\ct^{(1)} = \frac{r_1}{2}\left(\frac{2}{2+s} \right)^2=
 \left\{\begin{array}{ll} 
    0.0274(1)   &  \text{if $s=1$}, \\
   -0.08895(15) &  \text{if $s=0$},  \\
   -0.4636(6)   &  \text{if $s=-1$}, \\
 \end{array}\right.
\label{eq:ct1}
\end{equation}
The result for $s=0$ agrees within errors with the more precise value 
$\ct^{(1)}=-0.08900(5)$  quoted in ref.~\cite{Luscher:1993gh}.  
Note that $\ct$ is the coefficient of a local counterterm and thus cannot depend on
global space-time properties such as the aspect ratio $\rho$.
As a further check we also produced some data with $\rho=1/2$. 
Within the numerical accuracy, the coefficient $r_1$  is found to 
be proportional to $1/\rho$, which cancels the explicit factor 
$\rho$ in Eq.~(\ref{eq:ct1_formula}),
and renders $\ct^{(1)}$ $\rho$-independent, as expected.

\subsection{Residual cutoff effects in the step scaling function}

\begin{figure}[p]
\begin{center}
\begin{tabular}{c}
\includegraphics[width=0.80\textwidth]{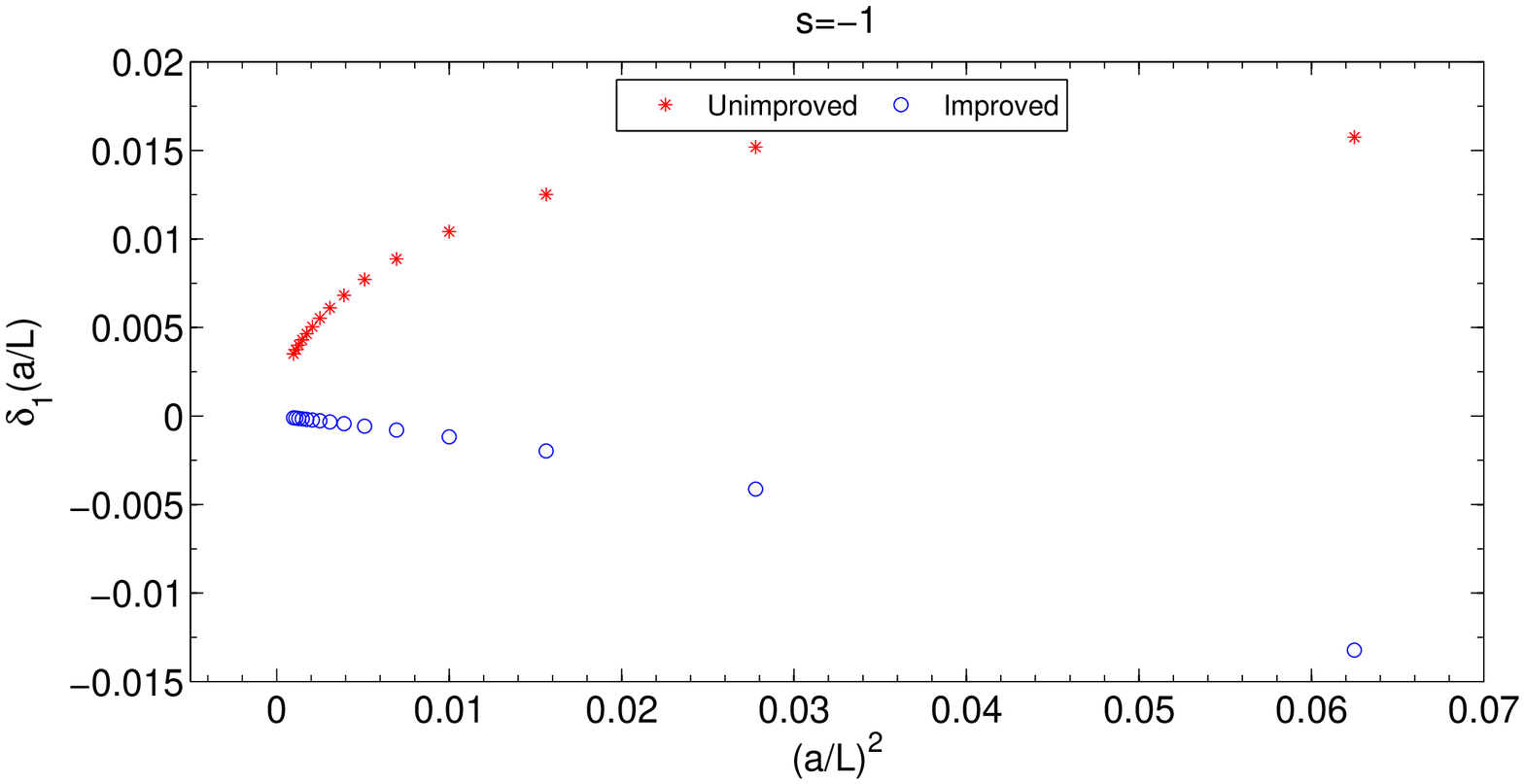}\\
\includegraphics[width=0.80\textwidth]{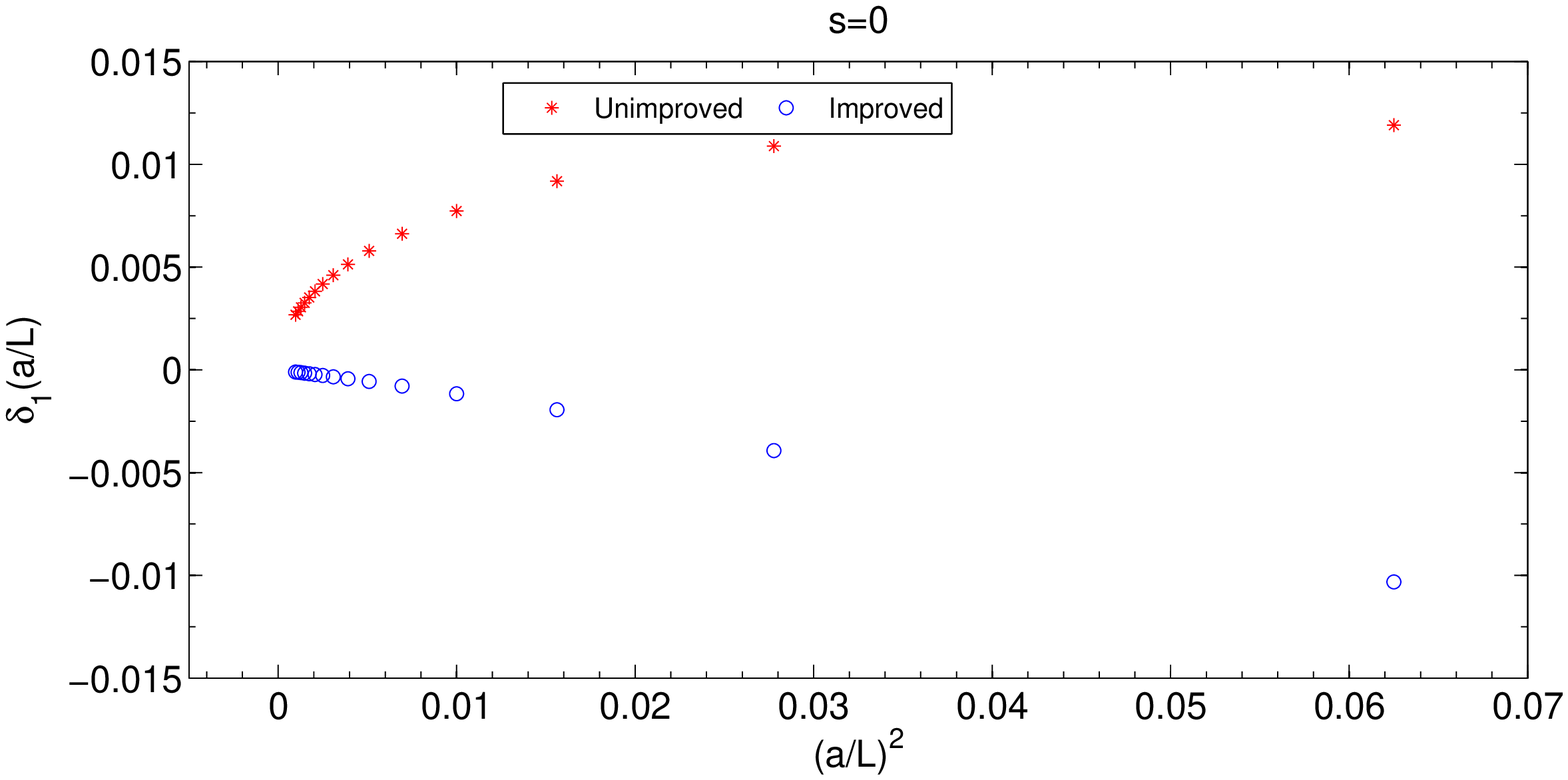}\\
\includegraphics[width=0.80\textwidth]{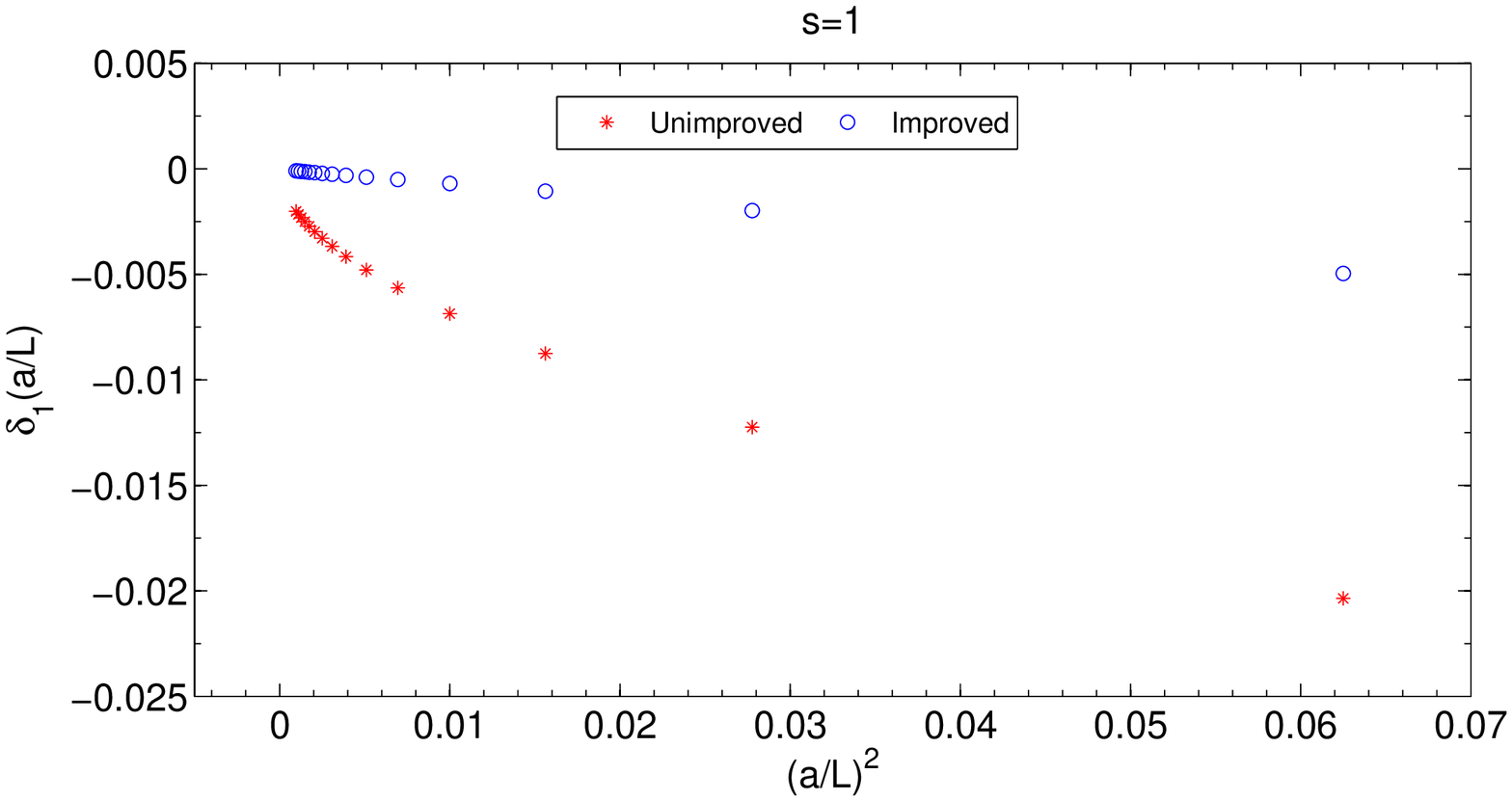}
\end{tabular}
\caption{\label{figure_delta1} Discretisation effects in the step-scaling function 
for $s =-1, 0,1$ from top to bottom, respectively.
Stars show the result obtained before cancellation of the ${\rm O}(a)$ part by the pure 
gauge boundary counterterm, $\ct^{(1)}$ and circles the results after the 
cancellation.}
\end{center}
\end{figure}

In the previous subsection we have removed the ${\rm O}(a)$ boundary lattice 
artefacts at one-loop order in perturbation theory. Nevertheless, higher 
order lattice artefacts are still present. We quantify these by   
studying the relative deviation of the lattice step scaling function, $\Sigma(u,a/L)$ 
from its continuum counterpart, $\sigma(u)$~[cf.~Eq.~(\ref{eq:SU3beta})], 
\begin{equation}
\delta(u,a/L) \equiv \frac{\Sigma(u,a/L) - \sigma(u)}{\sigma(u)}= \delta_1(a/L)u + \delta_2(a/L) u^2
 + {\rm O}(u^3).
\end{equation}
Defining 
\begin{equation}
   \overline{m}_1(L/a) = m_1(L/a) -\ct^{(1)}\left.\frac{\partial_\eta \Gamma_{0;\ct}[B]}{\partial_\eta\Gamma_0[B]}\right\vert_{\eta=0},
\end{equation}
a one-loop computation yields 
\begin{equation}
  \delta_1(a/L) = \overline{m}_1(2L/a) - \overline{m}_1(L/a) - 2b_0\ln(2),
\end{equation}
which we here use to monitor higher order lattice artefacts. In practice, $\delta_1(a/L)$
can also be used to cancel {\em all} lattice artefacts to O($u^2$) in non-perturbative
estimates of the step-scaling function. However, this implies that the $\ct$-counterterm must
be evaluated exactly and not just in its asymptotic form as in Eq.~(\ref{eq:Gam_ct_as}).
Using the field equation~(\ref{eq:varphi}), the resulting formulae are surprisingly 
simple, and one obtains,
\begin{equation}
   \frac{\partial_\eta \Gamma_{0;\ct}[B]}{\partial_\eta \Gamma_0[B]} =
  \frac{2}{\ct^{(0)}} \frac{a}{L}\left[
   \left(1+\frac{s}{2}-\rho\frac{L}{2a}\right)\partial_\eta\varphi +\frac{L}{2a}
                                \right] = \frac{a}{L} \frac{(2+s)^2}{2\rho} + \rmO(a^5).
\end{equation}
In the case $s=0$ one has $\partial_\eta \varphi \equiv 1/\rho$ from Eq.~(\ref{eq:varphi_0}) 
and the exact $\ct$-counterterm is thus found  to coincide with its asymptotic $a/L$-term. 
For $s=\pm 1$ the numerical difference is very small,
due to the mild cutoff effects in $\varphi$, Eq.~(\ref{eq:varphi_aexpansion}), and hence in $\partial_\eta\varphi$. 
In Table~\ref{table_delta1}  and in Fig.~\ref{figure_delta1}, 
we show two sets of data for $\delta_1(a/L)$, 
obtained by either setting $\ct^{(1)}=0$  or by choosing the 
correct value for $\ct^{(1)}$ from Eq.~(\ref{eq:ct1}).
Asymptotically one expects $\delta_1(a/L)$ to approach zero with a rate 
$\propto a/L$ and $\propto (a/L)^2$, respectively.
This is indeed observed in all cases. We also note that the data for all choices of the parameter $s$ 
seem to be behave similarly, i.e.~there is no clearly superior choice.
\begin{table}[ht!]
\begin{center}
\begin{tabular}{|l|l|l|l|l|l|l|}
\hline
     & \multicolumn{3}{c|}{$\delta_1(a/L)\vert_{\ct^{(1)}=0}$} & \multicolumn{3}{|c|}{$\delta_1(a/L)$} \\\cline{1-7} 
$L/a$& $s = 1$   & $s = 0$  & $s=-1$    & $s = 1$   & $s = 0$  & $s=-1$   \\\cline{1-7} 
$4$  & $-0.02036$& $0.01192$& $0.01575$ &$-0.00489$ &$-0.01033$&$ -0.01302$\\\cline{1-7} 
$6$  & $-0.01224$& $0.01089$& $0.01519$ &$-0.00195$ &$-0.00394$&$ -0.00409$\\\cline{1-7} 
$8$  & $-0.00876$& $0.00918$& $0.01252$ &$-0.00105$ &$-0.00194$&$ -0.00196$\\\cline{1-7} 
$10$ & $-0.00686$& $0.00773$& $0.01041$ &$-0.00069$ &$-0.00117$&$ -0.00117$\\\cline{1-7} 
$12$ & $-0.00564$& $0.00663$& $0.00887$ &$-0.00050$ &$-0.00079$&$ -0.00079$\\\cline{1-7} 
$14$ & $-0.00479$& $0.00579$& $0.00771$ &$-0.00039$ &$-0.00057$&$ -0.00057$\\\cline{1-7} 
$16$ & $-0.00416$& $0.00513$& $0.00682$ &$-0.00031$ &$-0.00043$&$ -0.00043$\\\cline{1-7} 
$18$ & $-0.00368$& $0.00461$& $0.00610$ &$-0.00025$ &$-0.00034$&$ -0.00033$\\\cline{1-7} 
$20$ & $-0.00329$& $0.00418$& $0.00553$ &$-0.00021$ &$-0.00027$&$ -0.00027$\\\cline{1-7} 
$22$ & $-0.00298$& $0.00382$& $0.00505$ &$-0.00018$ &$-0.00022$&$ -0.00022$\\\cline{1-7} 
$24$ & $-0.00272$& $0.00352$& $0.00464$ &$-0.00015$ &$-0.00019$&$ -0.00018$\\\cline{1-7} 
$26$ & $-0.00250$& $0.00326$& $0.00430$ &$-0.00013$ &$-0.00016$&$ -0.00016$\\\cline{1-7} 
$28$ & $-0.00232$& $0.00304$& $0.00400$ &$-0.00012$ &$-0.00013$&$ -0.00013$\\\cline{1-7} 
$30$ & $-0.00216$& $0.00285$& $0.00375$ &$-0.00010$ &$-0.00012$&$ -0.00012$\\\cline{1-7} 
$32$ & $-0.00202$& $0.00268$& $0.00352$ &$-0.00009$ &$-0.00010$&$ -0.00010$\\\cline{1-7} 
\hline
\end{tabular}

\caption{\label{table_delta1} One-loop coefficients $\delta_1(a/L)$
for $s= 1,0,-1$, without and with the one-loop counterterm $\propto \ct^{(1)}$.
For $s=\pm 1$ we have used Eq.~(\ref{eq:ct1}), for $s=0$ we set 
$\ct^{(1)}=-0.089$~\protect\cite{Luscher:1993gh}.}
\end{center}
\end{table} 

\subsection{Residual cutoff effects in $\bar v(L)$}

At fixed $u=\bar{g}^2(L)$ one may also study the lattice artefacts in the observable $\bar v(L)$ 
introduced in Section~2. On the lattice we define, 
\begin{equation}
\Omega(u,a/L) = \bar{v}(L)\vert_{u=\bar{g}^2(L)} = \Omega_1(a/L) + \Omega_2(a/L)u + \rmO(u^2),
\end{equation}
where $\Omega(u,a/L)$ converges to the universal function 
$\omega(u)$, Eq.~(\ref{eq:omega}) in the continuum limit.
In order to compute the one-loop coefficient $\Omega_1(a/L)$ it is convenient 
to consider a finite difference in $\nu$-values, rather than using Eq.~(\ref{eq:v_def}).
Both are equivalent since the $\nu$-dependence is completely explicit in Eq.~(\ref{eq:coupling}).
Hence,
\begin{equation}
    \bar{v}(L) = \frac{1}{(\nu_2 - \nu_1) \partial_\eta\Gamma_0[B]\vert_{\eta=0}}\left\{
   \left. \frac{\partial \Gamma[B]}{\partial \eta}\right|_{\eta=0,\nu=\nu_1}- 
   \left. \frac{\partial \Gamma[B]}{\partial \eta}\right|_{\eta=0,\nu=\nu_2} 
     \right\}.
  \label{eq:vbar_finitediff}
\end{equation}
Choosing $\nu_1=0$ and $\nu_2=1$ one obtains
\begin{equation}
 \Omega_1(a/L) =  m_1^{\nu=1}(L/a)-m_1(L/a),
 \label{eq:Omega1}
\end{equation}
with the one-loop coefficient [cf.~Eq.~(\ref{eq:hj})],
\begin{equation}
   m_1^{\nu=1} = \left[h_0 - \mbox{$\frac 12$}h_1\right]_{\nu=1}.
   \label{eq:m1nu1}
\end{equation}
Note that the $\ct$-counterterm is $\nu$-independent and does therefore not contribute in Eq.~(\ref{eq:Omega1}). 
We have computed $m_1^{\nu=1}$ for even lattice sizes up to $L/a=64$. 
By combining with the $\nu=0$ data we have checked that $\Omega_1(a/L)$ converges 
to the universal continuum value, $\omega_1$ of Eq.~(\ref{eq:vbar1loop}) for all cases $s=\pm 1,0$.  

To study the cutoff effects we follow ref.~\cite{Luscher:1993gh} and define,
\begin{equation}
  \epsilon(u,a/L)\equiv \frac{\Omega(u,a/L)-\omega(u)}{\omega(u)} = \epsilon_1(a/L) + \epsilon_2(a/L)u +\rmO(u^2). 
\end{equation}
At one-loop order we obtain
\begin{equation}
   \epsilon_1(a/L) = \Omega(a/L)/\omega_1 - 1,
\end{equation}
and we collect our numerical results in table~\ref{table_eps1}. The continuum 
limit is always approached with a rate of O($a^2$), as expected.
While the cutoff effects in $\Omega_1$ are sizeable, we do not observe a striking difference
between the cases $s=0,\pm 1$.

\begin{table}[ht!]
\begin{center}
\begin{tabular}{|l|l|l|l|}
\hline
$L/a$&  $s=-1$  &  $s=0$   & $s=1$     \\\cline{1-4} 
$4$  &$0.21001$&$0.18806$&$0.10209$\\\cline{1-4}
$6$  &$0.06874$&$0.06773$&$0.04449$\\\cline{1-4}
$8$  &$0.03365$&$0.03360$&$0.02419$\\\cline{1-4}
$10$ &$0.02040$&$0.02039$&$0.01562$\\\cline{1-4}
$12$ &$0.01380$&$0.01380$&$0.01104$\\\cline{1-4}
$14$ &$0.00999$&$0.00999$&$0.00825$\\\cline{1-4}
$16$ &$0.00758$&$0.00758$&$0.00641$\\\cline{1-4}
$18$ &$0.00595$&$0.00595$&$0.00513$\\\cline{1-4}
$20$ &$0.00480$&$0.00480$&$0.00420$\\\cline{1-4}
$22$ &$0.00395$&$0.00395$&$0.00350$\\\cline{1-4}
$24$ &$0.00331$&$0.00331$&$0.00297$\\\cline{1-4}
$26$ &$0.00282$&$0.00282$&$0.00254$\\\cline{1-4}
$28$ &$0.00243$&$0.00243$&$0.00221$\\\cline{1-4}
$30$ &$0.00211$&$0.00212$&$0.00193$\\\cline{1-4}
$32$ &$0.00185$&$0.00185$&$0.00171$\\\cline{1-4}
\hline
\end{tabular}

\caption{\label{table_eps1}One-loop coefficients $\epsilon_1(a/L)$
for $s=-1,0,1$}
\end{center}
\end{table}

  \section{A non-perturbative check}

In order to check our set-up non perturbatively, we have carried out numerical simulation
to compute the step scaling function and the observable $\bar v$. 
To generate a representative ensemble of gauge configurations, we used an
algorithm consisting in a combination of local heat bath sweeps (HB) with a number of
over-relaxation sweeps (OR). Measurements are performed for every cycle, where  
a cycle is defined by, 
\begin{figure}[htb]
\begin{center}
\includegraphics[width=0.8\textwidth]{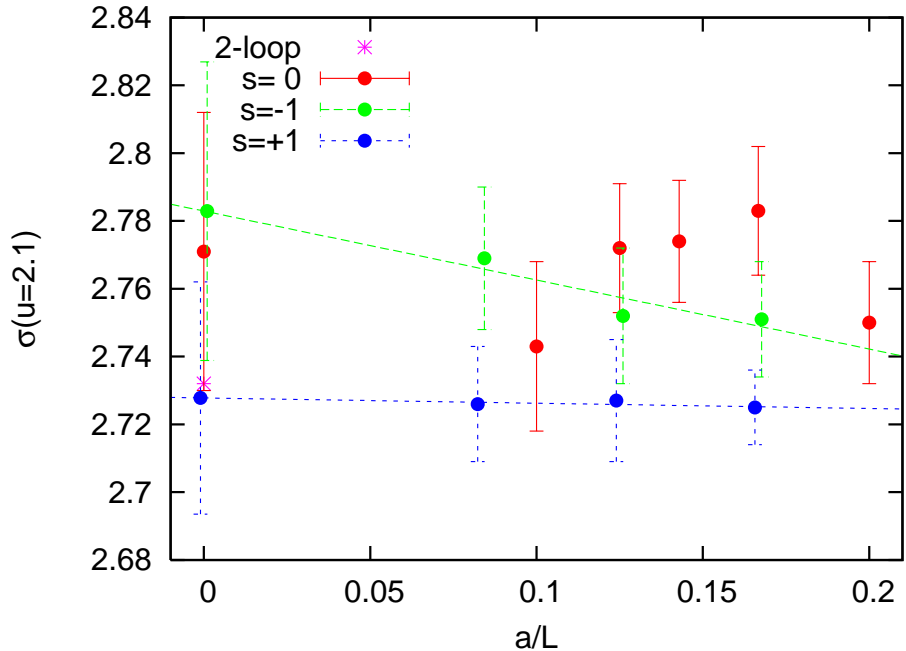}
\caption{\label{figure5}Cutoff dependence of the step 
scaling function at $u=2.1$ and continuum extrapolation. For comparison, results for $s=0$ taken from 
ref.~\protect\cite{Luscher:1993gh} are also shown.}
\end{center}
\end{figure}
\begin{equation}
1\,{\rm  cycle} = 1\,{\rm HB} + N_{\rm OR}\times {\rm OR},
\end{equation}
with $N_{\rm OR} = L/(2a)$. To study the evolution of the running 
coupling, pairs of lattices with sizes $L/a$ and $2L/a$ have been simulated. 
The results are shown in Table~\ref{table_simres}. In order to extract the 
mean value of the observables, as well as their uncertainties and their 
integrated autocorrelation times, we have used the procedure advocated in
ref.~\cite{Wolff:2003sm}. 

In Table~\ref{table_CL} we show the results for the step scaling function 
at fixed $u=2.1$. 
Figure~\ref{figure5} shows
the cutoff dependence of the step scaling function at $u=2.1$ and the 
continuum extrapolation. The results for $s=0$ from ref.~\cite{Luscher:1993gh}
are shown for comparison.

\begin{table}[ht!]
\begin{center}
\begin{tabular}{|l|l|l|l|l|l|l|l|}
\hline
 $s$ & $\beta$ &$L/a$ & $\bar g^2$ & $\tau_{\rm int}$ &
$\bar v$ & $\tau_{\rm int}$ &stat.  \\\cline{1-8} 
\multirow{2}{*}{$+1$}& \multirow{2}{*}{$7.2095$}&$6$&$2.0966(31)$& $0.64(1)$&$0.0579(15)$&
$1.49(4)$&$4700\times 16$ \\\cline{3-8}
    &        &$12$&$2.7191(96)$& $0.85(2)$&$0.0501(24)$&
$1.89(8)$&$4500\times 16$ \\\cline{1-8}
\multirow{2}{*}{$-1$}&\multirow{2}{*}{$7.1214$}&$6$&$2.0929(45)$& $0.58(1)$&$0.0610(16)$&
$0.91(2)$&$29700\times 4$ \\\cline{3-8}
    &        &$12$&$2.739(15) $& $0.64(2)$&$0.0470(29)$&
$0.98(4)$&$4700\times 16$ \\\cline{1-8}
\multirow{2}{*}{$+1$}&\multirow{2}{*}{$7.4218$}&$8$   &$2.0989(37)$& $0.68(1)$&$0.0602(17)$&
$1.69(6)$&$5500\times 16$ \\\cline{3-8}
 &    &$16$         &$2.725(17)$ & $0.93(4)$&$0.0524(37)$&
$1.74(10)$&$9500\times 4$ \\\cline{1-8}
\multirow{2}{*}{$-1$}&\multirow{2}{*}{$7.3632$}&$8$   &$2.0968(47)$& $0.58(1)$&$0.0587(16)$&
$0.93(2)$&$39700\times 4$ \\\cline{3-8}
 &    & $16$        &$2.747(18)$ & $0.59(2)$&$0.0557(36)$&
$0.95(4)$&$16700\times 4$ \\\cline{1-8}
\multirow{2}{*}{$+1$}&\multirow{2}{*}{$7.7447$}&$12$  &$2.0937(44)$& $0.77(2)$&$0.0578(20)$&
$1.90(6)$&$7500\times 6$ \\\cline{3-8}
 &    &$24$         &$2.715(15) $& $0.95(3)$&$0.0502(34)$&
$1.90(9)$&$19500\times 4$ \\\cline{1-8}
\multirow{2}{*}{$-1$}&\multirow{2}{*}{$7.6985$}&$12$  &$2.1017(55)$& $0.61(1)$&$0.0579(19)$&
$0.99(2)$&$49700\times4$ \\\cline{3-8}
 &    &$24$         &$2.772(19) $& $0.67(2)$&$0.0452(34)$&
$0.97(4)$&$29700\times 4$ \\\cline{1-8}
\hline
\end{tabular}

\caption{\label{table_simres} Simulation parameters and results for $\bar g^2$ 
and $\bar v$, and their autocorrelation time $\tau_{\rm int}$.}
\end{center}
\end{table} 

\begin{table}[htb]
\begin{center}
\begin{tabular}{|rl|c|}
\hline
\hline
$L/a$ & $s$ & $\sigma(u=2.1)$ \\\cline{1-3} 
$6$&$+1$ & $2.725(11)$\\
$6$&$-1$ & $2.751(17)$\\\cline{1-3}
$8$&$+1$ & $2.727(18)$\\
$8$&$-1$ & $2.752(21)$\\\cline{1-3}
$12$&$+1$& $2.726(17)$ \\
$12$&$-1$& $2.769(21)$\\\cline{1-3}
\hline
cont.&$+1$&$2.728(35)$  \\
cont.&$-1$&$2.783(44)$ \\\cline{1-3}
\hline
\end{tabular}

\caption{Continuum extrapolated results for the step scaling function at 
$u =2.1$ . Central values and error are corrected by following the strategy in
section~3.1 of ref.~\protect\cite{DellaMorte:2004bc}.}
\end{center}
\label{table_CL}
\end{table}

  \section{Conclusions}

Motivated by applications to lattice theories with staggered fermions, 
we have defined an O($a$) modified lattice set-up for the Schr\"odinger functional 
in the SU(3) gauge theory. Both perturbatively and non-perturbatively the cutoff
effects in the step-scaling function for the coupling are comparable to the standard set-up. 
Note that our approach can be applied to the SU(2) gauge theory with minor changes.
However, for SU($N$) with $N \ge 4$ this may or may not be the case, 
depending on the choices made for the boundary gauge fields. More precisely, assuming that
the induced lattice background field remains Abelian and spatially constant,
one may need to solve coupled equations in two or more variables replacing the single
equation for $\varphi$, Eq.~(\ref{eq:varphi}) of the SU(3) theory.

To extend this framework to four-flavour QCD and other QCD-like theories with staggered fermions,
one needs to discuss O($a$) boundary improvement with staggered fermions~\cite{PerezRubio:2008yd,PerezRubio:prep2}. 
While our motivation originates in applications using staggered fermions, we note
that there are other potential applications: for instance, the SF with Wilson 
or Ginsparg-Wilson fermions and chirally rotated boundary conditions is, for
technical convenience, constructed with an O($a$) offset in the orbifold 
construction~\cite{Sint:2010eh,Sint:2007zz}. This entails 
tree-level O($a$) artefacts in the fermionic propagator which could be avoided by 
keeping $T'/L$ fixed when taking the continuum limit, with $T'=T\pm a$. 
In any case, the new framework increases the flexibility for applications of the SF scheme.
Furthermore, it may help to improve control over the continuum limit, e.g.~by
performing constrained continuum extrapolations of data for step-scaling functions, computed 
at different values of the parameter $s$. This has already been attempted with staggered
fermions~\cite{PerezRubio:2010ke,PerezRubio:prep1} and it can be applied more generally.

  \subsection*{Acknowledgments}
We thank U.~Wolff for useful discussions. Much of this work was done while S.~T. was 
at the Humboldt University, partially supported by the SFB Transregio 9 of the DFG.
S.~S. acknowledges partial support by the EU under Grant Agreement number 
PITN-GA-2009-238353 (ITN STRONGnet).
P.~P\'erez-Rubio acknowledges support by the Spanish Ministery of Education 
and the DFG SFB/Transregio 55.

\newpage

\section*{A  Raw one-loop data}

%
%


\begin{table}[ht!]
\begin{center}
\begin{tabular}{|c|l|l|l|}
\hline
\hline
  $L/a$   &   $s = 1$       &     $s = 0$     &   $s =  -1$     \\\cline{1 - 4}
          $ 4 $  & $0.59867564635006385$ &  $0.52976131887602069$ &  $0.51922066331094546$ \\
          $ 6 $  & $0.64113765437070799$ &  $0.59321232599981455$ &  $0.58442701843014381$ \\
          $ 8 $  & $0.67488327354469054$ &  $0.63824441829314597$ &  $0.63153838271523941$ \\
          $ 10$  & $0.70238903864308802$ &  $0.67280038634701944$ &  $0.66742121983892173$ \\
          $ 12$  & $0.72546473348301643$ &  $0.70067317431899355$ &  $0.69618664881505825$ \\
          $ 14$  & $0.74530462655573632$ &  $0.72398165609402260$ &  $0.72013453722120221$ \\
          $ 16$  & $0.76269338777455524$ &  $0.74399271936483262$ &  $0.74062578372376441$ \\
          $ 18$  & $0.77816541775427566$ &  $0.76151561542170611$ &  $0.75852242009405113$ \\
          $ 20$  & $0.79209901867921951$ &  $0.77709645174594405$ &  $0.77440237357962767$ \\
          $ 22$  & $0.80477118528362614$ &  $0.79112035275707051$ &  $0.78867107103739184$ \\
          $ 24$  & $0.81639058776392414$ &  $0.80386875084304289$ &  $0.80162350146808351$ \\
          $ 26$  & $0.82711823835863092$ &  $0.81555339356121976$ &  $0.81348080743360567$ \\
          $ 28$  & $0.83708093056087357$ &  $0.82633751751530950$ &  $0.82441294090339149$ \\
          $ 30$  & $0.84638026935694360$ &  $0.83634957239642928$ &  $0.83455327855159342$ \\
          $ 32$  & $0.85509891481710033$ &  $0.84569242491756803$ &  $0.84400838361422786$ \\
          $ 34$  & $0.86330500808698590$ &  $0.85444971446735487$ &  $0.85286472295855835$ \\
          $ 36$  & $0.87105537841668160$ &  $0.86269035439316518$ &  $0.86119340950763814$ \\
          $ 38$  & $0.87839791234295844$ &  $0.87047179120264344$ &  $0.86905362653451155$ \\
          $ 40$  & $0.88537333426399287$ &  $0.87784241080234770$ &  $0.87649514951940602$ \\
          $ 42$  & $0.89201656536325291$ &  $0.88484334593899240$ &  $0.88356023620527862$ \\
          $ 44$  & $0.89835777515201167$ &  $0.89150985494698484$ &  $0.89028506544613721$ \\
          $ 46$  & $0.90442320537524014$ &  $0.89787238814607281$ &  $0.89670084803647420$ \\
          $ 48$  & $0.91023582290682148$ &  $0.90395742302994047$ &  $0.90283469523001835$ \\
          $ 50$  & $0.91581584250088787$ &  $0.90978812583979161$ &  $0.90871030564770146$ \\
          $ 52$  & $0.92118114931205926$ &  $0.91538488106299742$ &  $0.91434851426796011$ \\
          $ 54$  & $0.92634764338015540$ &  $0.92076571925410400$ &  $0.91976773540748381$ \\
          $ 56$  & $ 0.93132952274785759$  & $0.92594666572356148$  & $0.92498432332375374$ \\
          $ 58$  & $ 0.93613951787157274$  & $0.93094202701724371$  & $0.93001286814371605$ \\
          $ 60$  & $ 0.94078908704027068$  & $0.93576462803843164$  & $0.93486644054503752$ \\
          $ 62$  & $ 0.94528858033171669$  & $0.94042600966826691$  & $0.93955679547364682$ \\
          $ 64$  & $ 0.94964737799172085$  & $0.94493659452353015$  & $0.94409454285563946$ \\
\hline
\hline
\end{tabular}
\caption{\label{m1_data_even} The one-loop coefficient, $m_1(L/a)$, Eq.~(\ref{eq:m1}) 
for $s= 1,0,-1$. By comparing results of two independent calculations
we estimate that the number of significant digits decreases due to rounding errors from $15$ for $L/a=4$ 
to about $12$ for $L/a=64$.}
\end{center}
\end{table}

\begin{table}[ht!]
\begin{center}
\begin{tabular}{|c|l|l|l|}
\hline
\hline
  $L/a$   &   $s = 1$       &     $s = 0$     &   $s =  -1$     \\\cline{1 - 4}
         $ 5 $  & $0.62117136258700821$ &  $0.56468083125839683$ &  $0.55461713132090261$ \\
          $ 7 $  & $0.65893111142441375$ &  $0.61737811787009622$ &  $0.60974739393470162$ \\
          $ 9 $  & $0.68928647150162301$ &  $0.65654153428089878$ &  $0.65057003087364097$ \\
          $ 11$  & $0.71439398959413626$ &  $0.68741301652316919$ &  $0.68252028857956904$ \\
          $ 13$  & $0.73573276395961912$ &  $0.71280479111934452$ &  $0.70866242179288317$ \\
          $ 15$  & $0.75426748654695955$ &  $0.73434096296455847$ &  $0.73074988732724165$ \\
          $ 17$  & $0.77064247215528033$ &  $0.75302638176979129$ &  $0.74985728431314708$ \\
          $ 19$  & $0.78530529883645214$ &  $0.76952176876897567$ &  $0.76668599155622038$ \\
          $ 21$  & $0.79857842470262715$ &  $0.78428347860819714$ &  $0.78171762080518526$ \\
          $ 23$  & $0.81070148601313597$ &  $0.79763941632436668$ &  $0.79529658268786343$ \\
          $ 25$  & $0.82185727793917046$ &  $0.80983289063631291$ &  $0.80767742403345886$ \\
          $ 27$  & $0.83218834749448024$ &  $0.82104930186717609$ &  $0.81905346010774183$ \\
          $ 29$  & $0.84180796784671203$ &  $0.83143311025949667$ &  $0.82957488596710694$ \\
          $ 31$  & $0.85080762207365805$ &  $0.84109903085829275$ &  $0.83936067325760364$ \\
          $ 33$  & $0.85926224466335029$ &  $0.85013965596714345$ &  $0.84850663994188621$ \\
          $ 35$  & $0.86723397937992463$ &  $0.85863078810432712$ &  $0.85709107745685024$ \\
          $ 37$  & $0.87477492945292666$ &  $0.86663526032990589$ &  $0.86517876996699599$ \\
          $ 39$  & $0.88192920731676446$ &  $0.87420573028207099$ &  $0.87282392619177697$ \\
          $ 41$  & $0.88873448731299464$ &  $0.88138676139487858$ &  $0.88007235812526203$ \\
          $ 43$  & $0.89522319912792745$ &  $0.88821639863152864$ &  $0.88696312708410143$ \\
          $ 45$  & $0.90142345720374637$ &  $0.89472737905126752$ &  $0.89352980585123832$ \\
          $ 47$  & $0.90735979318226442$ &  $0.90094807414573837$ &  $0.89980145942895982$ \\
          $ 49$  & $0.91305373939615786$ &  $0.90690323216425240$ &  $0.90580341637673730$ \\
          $ 51$  & $0.91852429830862484$ &  $0.91261456924592534$ &  $0.91155788213294810$ \\
          $ 53$  & $0.92378832362922261$ &  $0.91810124483365024$ &  $0.91708443159609367$ \\
          $ 55$  & $ 0.92886083231152201$  & $0.92338024750431470$  & $0.92240040837837180$ \\
          $ 57$  & $0.93375526193982340$  & $0.92846671072289073$  & $0.92752125115677514$ \\
          $ 59$  & $ 0.93848368458072277$  & $0.93337417324448625$  & $0.93246076251996867$ \\
          $ 61$  & $0.94305698564346882 $  & $0.93811479540720059$  & $0.93723133204075556$ \\
          $ 63$  & $0.94748501439735390 $  & $0.94269953998217965$  & $0.94184412261958892$ \\
          $ 65$  & $ 0.95177671136930530$  & $0.94713832432409256$  & $0.94630922711111939$ \\
\hline
\hline
\end{tabular}
\caption{\label{m1_data_odd} Same as table~6 for lattices with odd $L/a$.}
\end{center}
\end{table}

\end{document}